\documentclass[aip,graphicx,reprint]{revtex4-1}

\usepackage{bm}
\usepackage{amsmath}
\usepackage{xcolor}
\usepackage{graphicx}
\usepackage{hyperref}
\usepackage[all]{hypcap}
\usepackage[mathlines]{lineno}

\draft

\makeatletter
\def\@email#1#2{%
 \endgroup
 \patchcmd{\titleblock@produce}
  {\frontmatter@RRAPformat}
  {\frontmatter@RRAPformat{\produce@RRAP{*#1\href{mailto:#2}{#2}}}\frontmatter@RRAPformat}
  {}{}
}%
\makeatother

\begin{document}
    \title{Linear damping of magneto-acoustic waves in two-fluid partially ionized plasmas}

    \author{David Martínez-Gómez}
    \affiliation{Departament de Física, Universitat de les Illes Balears, E-07122, Palma de Mallorca, Spain}
    \affiliation{Institut d'Aplicacions Computacionals de Codi Comunitari (IAC3), Universitat de les Illes Balears, E-07122, Palma de Mallorca, Spain}
    \email{david.martinez@uib.es}

    \date{\today}

    \begin{abstract}
        Magneto-acoustic waves in partially ionized plasmas are damped due to elastic collisions between charged and neutral particles. Here, we use a linearized two-fluid model to describe the influence of this collisional interaction on the properties of small-amplitude waves propagating in a uniform and static background. Mainly focusing on the case of waves generated by a periodic driver, we perform a detailed study of the dependence of the wavenumbers and damping rates on the ionization degree of the plasma, the strength of the collisional coupling, and the angle of propagation. We describe how the different wave modes (fast, slow, acoustic) are related to the individual properties of each fluid in a wide range of physical conditions. In addition, we derive analytical approximations for the damping rates due to charge-neutral collisions in the limits of weak and strong coupling and check their range of validity in comparison with the exact numerical results. These approximations can be generally applied to a large variety of astrophysical and laboratory partially ionized plasmas, but here we also discuss the particular application to plasmas only composed of hydrogen.
    \end{abstract}
    
    \maketitle

\section{Introduction} \label{sec:intro}
    Partially ionized plasmas consist of a mixture of electrically charged and neutral particles. They can be commonly found in many astrophysical environments such as the solar and planetary atmospheres or the interstellar medium \citep{Ballester2018SSRv..214...58B}, but also in laboratory scenarios such as experiments of magnetic reconnection \citep[see, e.g.,][]{Zweibel2011PhPl...18k1211Z,Lawrence2013PhRvL.110a5001L} or laser pulses \citep[see, e.g.,][]{Rowlands2008PhRvL.100j5005R}. Due to the fact that neutral particles are not directly affected by magnetic fields, their presence may have a strong impact in the properties of magnetohydrodynamic (MHD) waves \citep{Piddington1956MNRAS.116..314P,Kulsrud1969ApJ...156..445K}. Ionized and neutral particles interact by means of collisions, and one of the main effects of this interaction is the damping of MHD waves \citep{Watanabe1961CaJPh..39.1044W,Watanabe1961CaJPh..39.1197W,Braginskii1965RvPP....1..205B}, which eventually leads to heating of the plasma due to the dissipation of the energy contained in the waves \citep{Osterbrock1961ApJ...134..347O,Vayliunas2005JGRA..110.2301V,Leake2005A&A...442.1091L,Khomenko2012ApJ...747...87K,MartinezSykora2012ApJ...753..161M,Arber2016ApJ...817...94A}.

    Magneto-acoustic (or magneto-sonic) waves are MHD waves in which the compressibility of the plasma plays a relevant role, so they are driven by gas pressure forces in addition to the magnetic forces \citep{Lighthill1960RSPTA.252..397L,Goossens2003ASSL..294.....G,Goedbloed2004prma.book.....G}. The investigation of magneto-acoustic waves in partially ionized plasmas has been carried out both by means of single-fluid models in which a strong coupling between the electrically charged and neutral components of the plasma is assumed \citep[see, e.g.,][]{Forteza2007A&A...461..731F,Forteza2008A&A...492..223F,Soler2010A&A...512A..28S,Soler2015ApJ...810..146S} and by multi-fluid models in which charges and neutrals are treated as separate fluids \citep{Zaqarashvili2011A&A...529A..82Z,Mouschovias2011MNRAS.415.1751M,Soler2013ApJS..209...16S,Soler2024RSPTA.38230223S}. An overall conclusion from these studies is that the damping due to collisions between charges and neutrals is more efficient in the intermediate coupling regime (where the frequencies of the waves are of the same order of magnitude than the collision frequencies) and in plasmas with a low ionization fraction.

    Several works such as Refs. \onlinecite{KumarRoberts2003SoPh..214..241K,Forteza2007A&A...461..731F,Mouschovias2011MNRAS.415.1751M,Cally2023ApJ...954...85C,Soler2024RSPTA.38230223S,Molevich2024PhPl...31d2115M} have provided analytical expressions for the damping rates of small-amplitude waves generated by an impulsive driver. However, the study of propagating waves excited by a periodic driver has been commonly addressed either by numerically solving the dispersion equation that relates the frequencies of the waves with their wavenumbers or by performing numerical simulations \citep[see, e.g.,][]{Maneva2017ApJ...836..197M,Popescu2019A&A...627A..25P,Popescu2019A&A...630A..79P,Zhang2021ApJ...911..119Z,GomezMiguez2025A&A...701A.196G}, although some analytical results were derived for the particular case of hydrogen plasmas in Refs. \onlinecite{Carbonell2010A&A...515A..80C,CallyGomezMiguez2023ApJ...946..108C}. Thus, the main goals of the current investigation are to perform a comprehensive study of the solutions of the dispersion relation for periodically driven waves and to derive analytical approximations of the wavenumbers and spatial damping rates which can be applied to a general composition of the plasma within the two-fluid framework and which can help in interpreting the numerical results.

    In this work, a two-fluid model for partially ionized plasmas is used \citep[see, e.g.,][]{Meier2011PhDT.......208M,Leake2012ApJ...760..109L,Khomenko2014PhPl...21i2901K}, but restricting the investigation to the linear regime that describes the properties of small-amplitude magneto-acoustic waves \citep[][]{Soler2013ApJS..209...16S}. Therefore, non-linear features such as shocks \citep{Hillier2016A&A...591A.112H,Snow2021A&A...645A..81S} or plasma heating are not considered \citep[see, e.g.,][]{Soler2015ApJ...810..146S,Srivastava2021JGRA..12629097S}. In addition, only the interaction between charged and neutral particles by means of elastic collisions is taken into account, while other non-ideal processes such as ionization and recombination \citep{Ballai2019FrASS...6...39B}, Hall's current \citep{Amagishi1993PhRvL..71..360A,Pandey2008MNRAS.385.2269P}, or viscosity and thermal conduction \citep{Khodachenko2004A&A...422.1073K,Khodachenko2006AdSpR..37..447K,Forteza2008A&A...492..223F,Soler2010A&A...512A..28S,Barcelo2011A&A...525A..60B,PandeyWardle2024MNRAS.535.3410P} are neglected.

    This paper is organized as follows. Section \ref{sec:model} presents the basic definitions of the two-fluid model and the dispersion relation for compressible waves in a uniform background. Section \ref{sec:results} contains a detailed study of the solutions of the dispersion relation and the comparison between the analytical approximations and the exact numerical results, focusing on the damping rates of waves generated by a periodic driver. Section \ref{sec:discussion} includes a more general discussion of the obtained results, comparing them with those from previous works. Finally, Section \ref{sec:summary} summarizes the findings of the present research and details some possible improvements for future works.

\section{Model, basic equations and definitions} \label{sec:model}
    In the present study, the plasma is treated as a mixture of two fluids, in which one of the fluids, denoted by the subscript $c$, contains all the ionized or charged particles (including electrons) and the other fluid, denoted by the subscript $n$, contains all the electrically neutral particles. Then, the two fluids are allowed to interact by means of elastic collisions, but no other non-ideal effects, such as Hall's current, resistivity, viscosity, thermal conduction or inelastic collisions, are taken into account.

    Then, a static and uniform background is considered, in which the magnetic field is oriented along the $z$ direction, so $\bm{B}_{\rm{0}} = \left(0, 0, B_{\rm{0}} \right)$, and small-amplitude perturbations are applied to the background. The linearized version of the two-fluid equations that describe the evolution of those perturbations can be found, for instance, in Refs. \onlinecite{Zaqarashvili2011A&A...529A..82Z,Soler2013ApJ...767..171S,Soler2013ApJS..209...16S,Soler2024RSPTA.38230223S}.

    As shown in Ref. \onlinecite{Soler2013ApJS..209...16S}, assuming that the small-amplitude perturbations have a spatial dependence proportional to $\exp \left(i k_{x} x + i k_{y} y + i k_{z} z \right)$, where $k_{x}$, $k_{y}$, and $k_{z}$ are the components of the wavevector in the $x$, $y$, and $z$ directions, respectively, and a temporal dependence proportional to $\exp \left(-i \omega t \right)$, where $\omega$ is the wave frequency, the dispersion relation for two-fluid magnetoacoustic waves can be written as
    \begin{equation} \label{eq:ma_dr}
        \mathcal{D} \left(\omega, \bm{k}\right) = D_{\rm{c}}\left(\omega, \bm{k}\right) D_{\rm{n}}\left(\omega, \bm{k}\right) + D_{\rm{coll}}\left(\omega,\bm{k}\right) = 0,
    \end{equation}
    where
    \begin{gather} \label{eq:ma_drc}
        D_{\rm{c}}\left(\omega, \bm{k}\right) = \omega^{3} \left(\omega + i \nu_{\rm{cn}} \right) - \omega^{2} k^{2}\left(c_{\rm{A}}^{2} + c_{\rm{c}}^{2} \right) \nonumber \\
        +\frac{\omega + i \nu_{\rm{nc}}}{\omega+ i\left(\nu_{\rm{cn}} + \nu_{\rm{nc}}\right)}k^{4}c_{\rm{A}}^{2}c_{\rm{c}}^{2} \cos^{2} \theta,
    \end{gather}
    \begin{equation} \label{eq:ma_drn}
        D_{\rm{n}}\left(\omega, \bm{k} \right) = \omega \left(\omega + i \nu_{\rm{nc}} \right) - k^{2} c_{\rm{n}}^{2},
    \end{equation}
    and
    \begin{gather}
        D_{\rm{coll}} \left(\omega, \bm{k} \right) = \frac{\omega \nu_{\rm{cn}} \nu_{\rm{nc}}}{\omega + i \left(\nu_{\rm{cn}} + \nu_{\rm{nc}} \right)} \nonumber \\
        \times \left[\omega^{3} \Big(\omega + i \left(\nu_{\rm{cn}} + \nu_{\rm{nc}} \right) \Big) - k^{4} c_{\rm{A}}^{2} c_{\rm{n}}^{2} \cos^{2} \theta \right].
        \label{eq:ma_drcoll}
    \end{gather}
    
    The parameters $\nu_{\rm{cn}}$ and $\nu_{\rm{nc}}$ represent the charge-neutral and neutral-charge collision frequencies, $c_{\rm{A}}$ is the Alfvén speed of the charged fluid, $c_{\rm{c}}$ is the sound speed of the charged fluid, $c_{\rm{n}}$ is the sound speed of the neutral fluid, $k^{2} = k_{x}^{2} + k_{y}^{2} + k_{z}^{2}$ and $\theta$ is the angle formed by the wavevector $\bm{k}$ and the background magnetic field $\bm{B}_{\rm{0}}$.

    Equation (\ref{eq:ma_dr}) can be expanded in the following form:
    \begin{gather}  
        \Big[\Big(\omega^{4} + i \nu_{\rm{cn}} \omega^{3} - k^{2} \left(c_{\rm{A}}^{2} + c_{\rm{c}}^{2} \right) \omega^{2} \Big) \Big(\omega + i \left(\nu_{\rm{cn}} + \nu_{\rm{nc}}\right) \Big) \nonumber \\ 
        + k^{2} k_{z}^{2} c_{\rm{A}}^{2} c_{\rm{c}}^{2} \left(\omega + i \nu_{\rm{nc}}\right) \Big] \Big[\omega^{2} - k^{2} c_{\rm{n}}^{2} + i \nu_{\rm{nc}} \omega \Big] \nonumber \\
        + \nu_{\rm{cn}} \nu_{\rm{nc}} \omega \Big[ \omega^{3} \Big(\omega + i \left(\nu_{\rm{cn}} + \nu_{\rm{nc}} \right) \Big) - k^{2} k_{z}^{2} c_{\rm{A}}^{2} c_{\rm{n}}^{2} \Big] = 0
        \label{eq:ma_dr_full},
    \end{gather}
    where the relation $k_{z} = k \cos \theta$ has been applied. This is the same expression, although with a slightly different notation, as the one presented in Refs. \onlinecite{Zaqarashvili2011A&A...529A..82Z,Ballester2018SSRv..214...58B}.

    The sound speeds of the charged and the neutral fluids are given by
    \begin{equation}
        c_{\rm{c}} = \sqrt{\frac{\gamma P_{\rm{c0}}}{\rho_{\rm{c0}}}} \quad \text{and} \quad c_{\rm{n}} = \sqrt{\frac{\gamma P_{\rm{n0}}}{\rho_{\rm{n0}}}},
    \end{equation}
    where $P_{\rm{c0}}$ and $P_{\rm{n0}}$ are the background pressures of the charged and neutral fluids, respectively, $\rho_{\rm{c0}}$ and $\rho_{\rm{n0}}$ are the corresponding background densities, and $\gamma = 5/3$ is the adiabatic constant for a monatomic gas.

    The Alfvén speed of the charged fluid is given by
    \begin{equation}
        c_{\rm{A}} = \frac{B_{\rm{0}}}{\sqrt{\mu_{\rm{0}} \rho_{\rm{c0}}}},
    \end{equation}
    where $\mu_{\rm{0}}$ is the vacuum magnetic permeability.

    For later use, it is convenient to define the neutral-to-charge density ratio as $\chi = \rho_{\rm{n0}} / \rho_{\rm{c0}}$ and the global or effective Alfvén and sound speeds (which take into account the contributions from all the components of the plasma) as
    \begin{equation} \label{eq:calf_tot}
        a = \frac{c_{\rm{A}}}{\sqrt{1 + \chi}}
    \end{equation}
    and
    \begin{equation} \label{eq:ceff}
        c = \sqrt{\frac{\rho_{\rm{c0}} c_{\rm{c}}^{2} + \rho_{\rm{n0}} c_{\rm{n}}^{2}}{\rho_{\rm{c0}} + \rho_{\rm{n0}}}}  = \sqrt{\frac{c_{\rm{c}}^{2} + \chi c_{\rm{n}}^{2}}{1 + \chi}},
    \end{equation}
    which clearly fulfills that $c = c_{\rm{c}}$ if $\chi \to 0$ and $c = c_{\rm{n}}$ if $\chi \to \infty$. In addition, due to the momentum conservation, the collision frequencies fulfill that $\rho_{\rm{c0}} \nu_{\rm{cn}} = \rho_{\rm{n0}} \nu_{\rm{nc}}$, which leads to the relation $\nu_{\rm{cn}} = \chi \nu_{\rm{nc}}.$

\section{Results} \label{sec:results}
    The basic properties of magneto-acoustic waves in two-fluid partially ionized plasmas can be extracted from the dispersion relation given by Eqs. (\ref{eq:ma_dr}) or (\ref{eq:ma_dr_full}). The real part of the solutions of these equations provide the oscillation frequency or the wavenumber of the perturbations, depending on whether waves are generated by an impulsive or by a periodic driver. Then, the imaginary part provides the damping rate due to the collisional interaction between the two-fluids. Due to the complexity of the dispersion relation, the study of these properties for a wide range of physical conditions (such as different ionization degrees, propagation angles or coupling degrees) typically requires a numerical approach. However, some analytical approximations can be derived for certain scenarios.
    
\subsection{Magneto-acoustic waves: perpendicular propagation} \label{sec:ma_perp}
    In the first place, we study the case of waves propagating along the perpendicular direction to the background magnetic field. Therefore, setting $k_{z} = 0$ or, equivalently, $\theta = \pm \pi/2$, Eq. (\ref{eq:ma_dr_full}) simplifies to
    \begin{gather}  
        \omega^{2} \Big(\omega + i \left(\nu_{\rm{cn}} + \nu_{\rm{nc}}\right) \Big) \Big[\Big(\omega^{2} + i \nu_{\rm{cn}} \omega - k^{2} \left(c_{\rm{A}}^{2} + c_{\rm{c}}^{2} \right) \Big)  \nonumber \\ 
        \times \Big(\omega^{2} - k^{2} c_{\rm{n}}^{2} + i \nu_{\rm{nc}} \omega \Big) + \nu_{\rm{cn}} \nu_{\rm{nc}} \omega^{2} \Big] = 0.
        \label{eq:ma_dr_perp}
    \end{gather}
    It can be seen that this equation is a $7^{th}$ order polynomial in $\omega$ but a $4^{th}$ order polynomial in $k$, so it will lead to a different number of oscillation modes depending on whether an impulsive or a periodic driver is considered. In addition, Eq. (\ref{eq:ma_dr_perp}) can be split in three clearly different branches. The first two branches are functions of the wave frequency but not of the wavenumber. They provide the following solutions:
    \begin{equation}
        \omega^{2} = 0,
    \end{equation}
    corresponding to the slow magneto-acoustic waves, which do not propagate along the perpendicular direction to the background magnetic field, and
    \begin{equation} \label{eq:w_flowdiff}
         \omega = -i \left(\nu_{\rm{nc}} + \nu_{\rm{cn}} \right) = -i \left(1 + \chi \right) \nu_{\rm{nc}},
    \end{equation}
    representing an evanescent (non-propagating) mode, which has been denoted as \textit{flow differential mode} in Ref. \onlinecite{Cally2023ApJ...954...85C} due to its relation with the drift velocity between the two fluids.
    
    In the remainder of this section we focus on the third branch from Eq. (\ref{eq:ma_dr_perp}), which depends both in $\omega$ and $k$, and contains a combination of the fast magneto-acoustic waves of the charged fluid and the acoustic waves of the neutral fluid.
    
\subsubsection{Weak coupling regime for perpendicular propagation} \label{sec:ma_perp_weak}
    If the wave frequency is much larger than the collision frequencies, that is, if $\omega \gg \{\nu_{\rm{cn}}, \nu_{\rm{nc}} \}$, the last term in Eq. (\ref{eq:ma_dr_perp}) can be neglected, and we obtain two independent dispersion relations, one for each fluid. The case of standing waves has already been described in Ref. \onlinecite{Soler2024RSPTA.38230223S}, so here we concentrate on the case of propagating waves with $\omega$ real and $k$ complex.

    The dispersion relation associated with the charged fluid is given by
    \begin{equation} \label{eq:k2_perp_weak_fast}
        k^{2} = \frac{\omega^{2}}{c_{\rm{A}}^{2} + c_{\rm{c}}^{2}} + \frac{i \omega \nu_{\rm{cn}}}{c_{\rm{A}}^{2} + c_{\rm{c}}^{2}}.
    \end{equation}
    Assuming that $k = k_{\rm{R}} + i k_{\rm{I}}$, with $k_{\rm{I}} \ll k_{\rm{R}}$ (so the damping is weak), the approximate solution to Eq. (\ref{eq:k2_perp_weak_fast}) is
    \begin{equation} \label{eq:ksol_perp_weak_fast}
        k \approx \pm \left(\frac{\omega}{\sqrt{c_{\rm{A}}^{2} + c_{\rm{c}}^{2}}} + i\frac{\nu_{\rm{cn}}}{2 \sqrt{c_{\rm{A}}^{2} + c_{\rm{c}}^{2}}}\right),
    \end{equation}
    where the plus sign corresponds to forward propagating waves and the minus sign corresponds to backward propagating waves. We see that the damping rate ($k_{\rm{I}}$) of this mode does not depend on the oscillation frequency but it is proportional to the charge-neutral collision frequency, $\nu_{\rm{cn}}$.

    Similarly, the dispersion relation associated with the neutral fluid is
    \begin{equation} \label{eq:k2_perp_weak_sound}
        k^{2} = \frac{\omega^{2}}{c_{\rm{n}}^{2}} + \frac{i \omega \nu_{\rm{nc}}}{c_{\rm{n}}^{2}},
    \end{equation}
    whose approximate solution is
    \begin{equation} \label{eq:ksol_perp_weak_sound}
        k \approx \pm \left(\frac{\omega}{c_{\rm{n}}} + i \frac{\nu_{\rm{nc}}}{2 c_{\rm{n}}} \right).
    \end{equation}
    The damping rate of this mode is proportional to the neutral-charge collision frequency, $\nu_{\rm{nc}}$.

    In addition, Eqs. (\ref{eq:ksol_perp_weak_fast}) and (\ref{eq:ksol_perp_weak_sound}) show that the damping rates of the fast and neutral-acoustic modes are inversely proportional to their respective phase speeds. 

\subsubsection{Strong coupling regime for perpendicular propagation} \label{sec:ma_perp_coupled}
    Here, we consider the scenario in which the two fluids that compose the plasma are strongly coupled by collisions, with $\omega \ll \{\nu_{\rm{cn}}, \nu_{\rm{nc}} \}$. To analyze this case it is useful to write the dispersion relation in the following form \citep[see, e.g.,][]{Popescu2019A&A...630A..79P}: 
    \begin{gather}
        \Big(\omega^{2} - k^{2} \left(c_{\rm{A}}^{2} + c_{\rm{c}}^{2} \right) \Big) \Big(\omega^{2} - k^{2} c_{\rm{n}}^{2} \Big) +i \nu_{\rm{cn}} \omega \Big(\omega^{2} - k^{2} c_{\rm{n}}^{2} \Big) \nonumber \\
         + i \nu_{\rm{nc}} \omega \Big(\omega^{2} - k^{2} \left(c_{\rm{A}}^{2} + c_{\rm{c}}^{2} \right)\Big) = 0
        \label{eq:ma_perp}
    \end{gather}
    or
    \begin{equation} \label{eq:dr_perp_ds}
        \mathcal{D}^{\perp}\left(\omega, \bm{k} \right) = D_{\rm{c}}^{\perp}\left(\omega, \bm{k} \right) D_{\rm{n}}^{\perp}\left(\omega, \bm{k} \right) + D_{\rm{coll}}^{\perp}\left(\omega, \bm{k} \right) = 0,
    \end{equation}
    where
    \begin{equation}
        D_{\rm{c}}^{\perp}\left(\omega, \bm{k} \right) = \omega^{2} - k^{2} \left(c_{\rm{A}}^{2}+ c_{\rm{c}}^{2} \right),
    \end{equation}
    \begin{equation}
        D_{\rm{n}}^{\perp}\left(\omega, \bm{k} \right) = \omega^{2} - k^{2} c_{\rm{n}}^{2},
    \end{equation}
    and
     \begin{equation}
        D_{\rm{coll}}^{\perp}\left(\omega, \bm{k} \right) = i \nu_{\rm{cn}} \omega D_{\rm{n}}^{\perp}\left(\omega, \bm{k} \right) + i \nu_{\rm{nc}} \omega D_{\rm{c}}^{\perp}\left(\omega, \bm{k} \right).
    \end{equation}
    
    Since the collision frequencies are now assumed to be much larger than the wave frequency, we have that $\mathcal{D}^{\perp}\left(\omega, \bm{k} \right) \approx D_{\rm{coll}}^{\perp} \left(\omega, \bm{k} \right)$, and the term $D_{\rm{c}}^{\perp}\left(\omega, \bm{k} \right) D_{\rm{n}}^{\perp}\left(\omega, \bm{k} \right)$ can be considered as a small correction. Then, we follow the method detailed in Ref. \onlinecite{Tagger1995A&A...299..940T} to compute the solutions for the case of weak damping and write the dispersion relation in the form $\mathcal{D}^{\perp}(\omega,k) = D_{\rm{R}}^{\perp} + i D_{\rm{I}}^{\perp}$, with
    \begin{gather}
        D_{\rm{R}}^{\perp} = -i D_{\rm{coll}}^{\perp}\left(\omega, \bm{k} \right) = \nonumber \\ \nu_{\rm{nc}} \omega \left[\left(1 + \chi \right) \omega^{2} - k^{2} \left(c_{\rm{A}}^{2} + c_{\rm{c}}^{2} + \chi c_{\rm{n}}^{2} \right) \right]
    \end{gather}
    and
    \begin{equation}
        D_{\rm{1}}^{\perp} = -i D_{\rm{c}}^{\perp}\left(\omega, \bm{k} \right) D_{\rm{n}}^{\perp}\left(\omega, \bm{k} \right).
    \end{equation}

    In this way, from the equation $D_{\rm{R}}^{\perp} = 0$ we can obtain the real part of the solutions. For waves generated by a periodic driver, with $\omega$ real and $k = k_{\rm{R}} + i k_{\rm{I}}$, we have that
    \begin{equation} \label{eq:perp_strong_krd0}
        k_{\rm{R}} \approx \pm \sqrt{\frac{1 + \chi}{c_{\rm{A}}^{2} + c_{\rm{c}}^{2} + \chi c_{\rm{n}}^{2}}} \omega = \pm \frac{\omega}{\sqrt{a^{2} + c^{2}}}. 
    \end{equation}
    For waves generated by an impulsive driver, with $k$ real and $\omega = \omega_{\rm{R}} + i \omega_{\rm{I}}$, we find an equivalent expression to Eq. (\ref{eq:perp_strong_krd0}) but making the substitutions $k_{\rm{R}} \to k$ and $\omega \to \omega_{\rm{R}}$, and an additional solution with
    \begin{equation} \label{eq:wr0}
        \omega_{\rm{R}} = 0.
    \end{equation}

    Equation (\ref{eq:perp_strong_krd0}) gives the wavenumber of the global or modified fast waves \citep[see, e.g.,][]{Soler2013ApJS..209...16S}, which takes into account the contributions from both the charged and neutral fluids.

    The damping rates for the cases with $\omega$ real and $k$ real can be computed as
    \begin{equation} \label{eq:ki_wi_tagger}
        k_{\rm{I}} = -\frac{D_{\rm{I}}^{\perp}}{\partial D_{\rm{R}}^{\perp}/\partial k} \Bigg|_{\omega,k_{\rm{R}}} \quad \text{and} \quad \omega_{\rm{I}} = -\frac{D_{\rm{I}}^{\perp}}{\partial D_{\rm{R}}^{\perp}/\partial \omega} \Bigg|_{k,\omega_{\rm{R}}},
    \end{equation}
    respectively. This procedure results in the following expressions:
    \begin{equation} \label{eq:perp_strong_kId1}
        k_{\rm{I}} = \pm \frac{\chi \left(c_{\rm{A}}^{2} + c_{\rm{c}}^{2} - c_{\rm{n}}^{2} \right)^{2}\omega^{2}}{2 \nu_{\rm{nc}} \sqrt{1 + \chi} \left(c_{\rm{A}}^{2} + c_{\rm{c}}^{2} +\chi c_{\rm{n}}^{2} \right)^{5/2}},
    \end{equation}
    and
    \begin{equation} \label{eq:perp_strong_wId1}
        \omega_{\rm{I}} =  - \frac{\chi \left(c_{\rm{A}}^{2} + c_{\rm{c}}^{2} - c_{\rm{n}}^{2} \right)^{2}k^{2}}{2 \nu_{\rm{nc}} \left(1 + \chi \right)^{2} \left(c_{\rm{A}}^{2} + c_{\rm{c}}^{2} +\chi c_{\rm{n}}^{2} \right)},
    \end{equation}
    which show that the damping rates increase with the square of the oscillation frequency or with the square of the wavenumber, and that they are inversely proportional to the collision frequency. Therefore, the damping rates decrease as the two fluids become more strongly coupled, as it has been usually found for MHD waves in partially ionized plasmas \citep[see, e.g.,][]{Braginskii1965RvPP....1..205B,Forteza2007A&A...461..731F,Zaqarashvili2011A&A...529A..82Z,Soler2013ApJ...767..171S,Ballester2018SSRv..214...58B}.
    
    Up to now, we have obtained three solutions for $\omega$ and two solutions for $k$. However, the dispersion relation given by Eq. (\ref{eq:ma_perp}) is of fourth order in both parameters. Therefore, we are still missing one solution for $\omega$ and two for $k$. To find these additional solutions we can take advantage of the fact that Eq. (\ref{eq:ma_perp}) is a bi-quadratic equation in $k$, so it fulfills that
    \begin{gather} 
        \left(k^{2} - r_{\rm{1}} \right) \left(k^{2} - r_{\rm{2}} \right) =0 \nonumber \\
        \Rightarrow k^{4} -\left(r_{\rm{1}} + r_{\rm{2}} \right)k^{2} + r_{\rm{1}} r_{\rm{2}} = 0,
        \label{eq:biquad}
    \end{gather}
    where $r_{\rm{1}}$ and $r_{\rm{2}}$ are the roots of the bi-quadratic equation. If we rearrange Eq. (\ref{eq:ma_perp}) and compare it to Eq. (\ref{eq:biquad}), we obtain the corresponding version of Vieta's formulas:
    \begin{equation} \label{eq:vieta1}
        r_{\rm{1}} + r_{\rm{2}} = \frac{\mathcal{F}(\omega)}{\left(c_{\rm{A}}^{2} + c_{\rm{c}}^{2} \right) c_{\rm{n}}^{2}},
    \end{equation}
    where
    \begin{gather}
        \mathcal{F}(\omega) = \left(c_{\rm{A}}^{2} + c_{\rm{c}}^{2} + c_{\rm{n}}^{2} \right) \omega^{2} \nonumber \\
        + i \chi \nu_{\rm{nc}} \omega c_{\rm{n}}^{2} + i \nu_{\rm{nc}} \omega \left(c_{\rm{A}}^{2} + c_{\rm{c}}^{2} \right)
    \end{gather}
    and
    \begin{equation} \label{eq:vieta2}
        r_{\rm{1}} r_{\rm{2}} = \frac{\omega^{3} \left(\omega + i \left(1 + \chi \right) \nu_{\rm{nc}} \right)}{\left(c_{\rm{A}}^{2} + c_{\rm{c}}^{2} \right) c_{\rm{n}}^{2}}.
    \end{equation}
    If we assume that one the roots of the bi-quadratic equation is approximately given by the real part of the global fast wave, so $r_{\rm{1}} \approx k_{\rm{R}}^{2}$, with $k_{\rm{R}}$ given by Eq. (\ref{eq:perp_strong_krd0}), the root $r_{\rm{2}}$ can be obtained by solving either of the two Vieta's relations. From Eq. (\ref{eq:vieta1}) we find that
    \begin{align} \label{eq:r2_perp_1}
        r_{\rm{2}} = k_{\rm{2}}^{2} \approx \frac{i \nu_{\rm{nc}} \omega \left(c_{\rm{A}}^{2} + c_{\rm{c}}^{2} + \chi c_{\rm{n}}^{2} \right)}{\left(c_{\rm{A}}^{2} + c_{\rm{c}}^{2} \right) c_{\rm{n}}^{2}} \nonumber \\
        + \frac{\left(c_{\rm{A}}^{4} + 2 c_{\rm{A}}^{2} c_{\rm{c}}^{2} + c_{\rm{c}}^{4} + \chi c_{\rm{n}}^{4} \right) \omega^{2}}{\left(c_{\rm{A}}^{2} + c_{\rm{c}}^{2} \right) c_{\rm{n}}^{2} \left(c_{\rm{A}}^{2} + c_{\rm{c}}^{2} + \chi c_{\rm{n}}^{2} \right)},
    \end{align}
    and from Eq. (\ref{eq:vieta2}) we get
    \begin{equation} \label{eq:r2_perp_2}
        r_{\rm{2}} = k_{\rm{2}}^{2} \approx \frac{\omega \left(\omega + i \left(1 + \chi \right) \nu_{\rm{nc}} \right) \left(c_{\rm{A}}^{2} + c_{\rm{c}}^{2} + \chi c_{\rm{n}}^{2} \right)}{\left(c_{\rm{A}}^{2} + c_{\rm{c}}^{2} \right) c_{\rm{n}}^{2} \left(1 + \chi \right)}.
    \end{equation}
    We have obtained two different approximations. However, in the limit of interest for this section (when $\omega \ll \nu_{\rm{nc}}$) both reduce to the expression
    \begin{equation} \label{eq:sols_perp_strong2}
        k_{\rm{2}}^{2} \approx \frac{i \nu_{\rm{nc}} \omega \left( c_{\rm{A}}^{2} + c_{\rm{c}}^{2} + \chi c_{\rm{n}}^{2} \right)}{\left(c_{\rm{A}}^{2} + c_{\rm{c}}^{2} \right) c_{\rm{n}}^{2}},
    \end{equation}
    showing that the square of the wavenumber is a purely imaginary number. Therefore, the real part and the imaginary part of the wavenumber will have the same value, which will be proportional to the square root of the oscillation frequency and of the collision frequency.

    Note that we have obtained Eq. (\ref{eq:sols_perp_strong2}) by assuming that $\omega$ is real and $k$ complex. However, it can be shown that an identical expression results from assuming a real $k$ and a complex $\omega$ and applying the respective Vieta's formulas to a polynomial of the form $\sum_{\rm{i = 0}}^{4} A_{\rm{i}} \omega^{\rm{i}} = 0$, where $A_{\rm{i}}$ are functions of $k$, the collision frequencies and the characteristic speeds. In that case, we find a mode with a negative purely imaginary wave frequency which is inversely proportional to the collisional frequency $\nu_{\rm{nc}}$. In addition, we find that the imaginary part of the mode with $\omega_{\rm{R}} = 0$ mentioned in previous paragraphs is $\omega_{\rm{I}} = -i \left(1 + \chi \right) \nu_{\rm{nc}}$, as in Eq. (\ref{eq:w_flowdiff}). Therefore, in the limit of strong collisional coupling, this expression appears as a double root of the dispersion relation, in agreement with the findings of Refs. \onlinecite{Mouschovias2011MNRAS.415.1751M,Soler2024RSPTA.38230223S}.

    Once the expressions for the approximate solutions of the dispersion relation have been derived, it is also interesting to check their behaviour as functions of the ionization degree of the plasma. In the limit of strongly ionized plasma, with $\chi \to 0$, the real part of the wavenumber of the global fast mode, given by Eq. (\ref{eq:perp_strong_krd0}), reduces to $k_{\rm{R}} \approx \pm \omega / \sqrt{c_{\rm{A}}^{2} + c_{\rm{c}}^{2}}$, so it can be mainly associated with the fast mode of the charged fluid alone, while the damping rates given by Eqs. (\ref{eq:perp_strong_kId1}) and (\ref{eq:perp_strong_wId1}) will tend to $0$. In addition, Eq. (\ref{eq:sols_perp_strong2}) becomes $k_{\rm{2}}^{2} \approx i \nu_{\rm{nc}} \omega / c_{\rm{n}}^{2}$, representing a strongly damped neutral acoustic mode. On the other hand, if $\chi c_{\rm{n}}^{2} \gg c_{\rm{A}}^{2} + c_{\rm{c}}^{2}$, corresponding to the limit of weakly ionized plasmas, the global fast mode will be an almost undamped neutral acoustic mode, with $k_{\rm{R}} \approx \pm \omega / c_{\rm{n}}$, while the strongly attenuated mode can be identified with the fast wave of the charged fluid, since Eq. (\ref{eq:sols_perp_strong2}) now becomes $k_{\rm{2}}^{2} \approx i \nu_{\rm{cn}} \omega / \left(c_{\rm{A}}^{2} + c_{\rm{c}}^{2} \right)$.

\subsubsection{Parametric study and comparison between analytical approximations and exact numerical solutions of the dispersion relation} \label{sec:ma_perp_general}
    \begin{figure*}
        \centering
        \includegraphics[width=0.49\hsize]{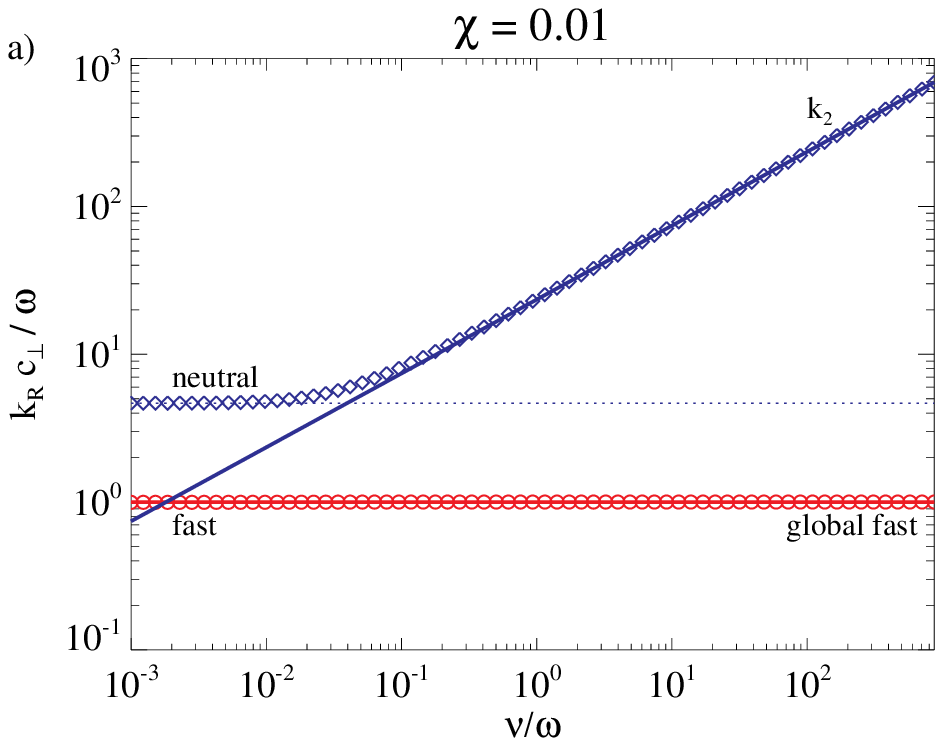}
        \includegraphics[width=0.49\hsize]{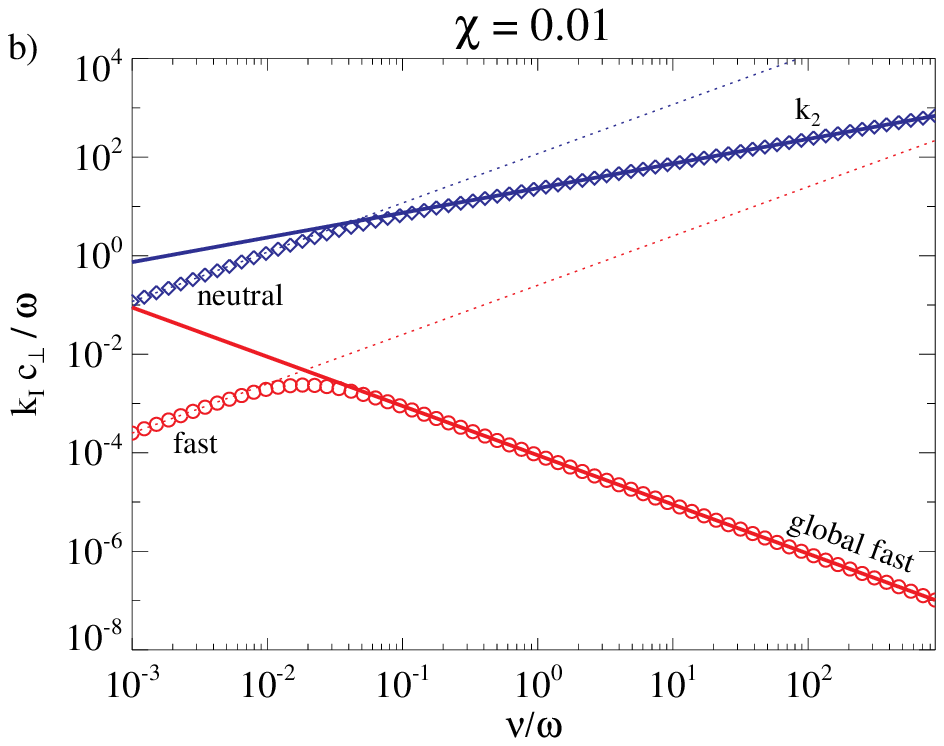} \\
        \includegraphics[width=0.49\hsize]{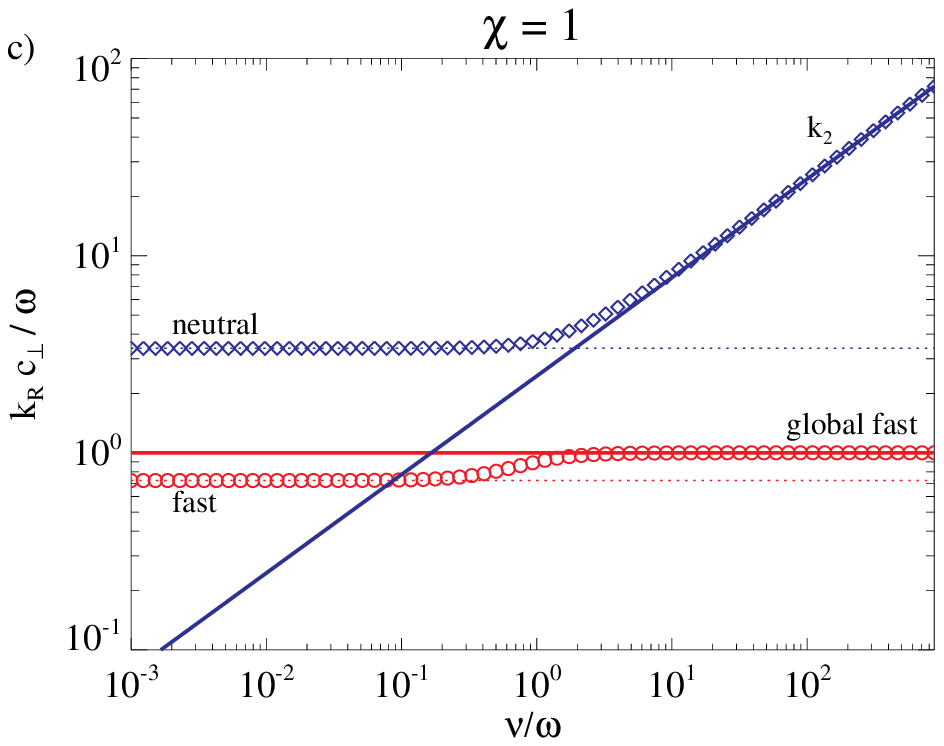}
        \includegraphics[width=0.49\hsize]{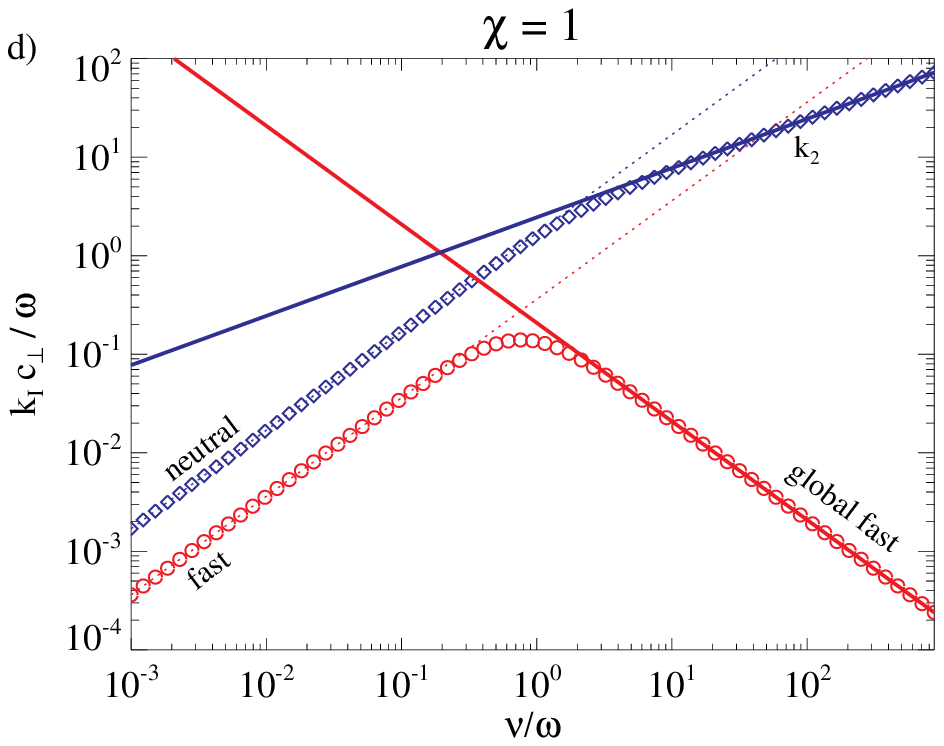} \\
        \includegraphics[width=0.49\hsize]{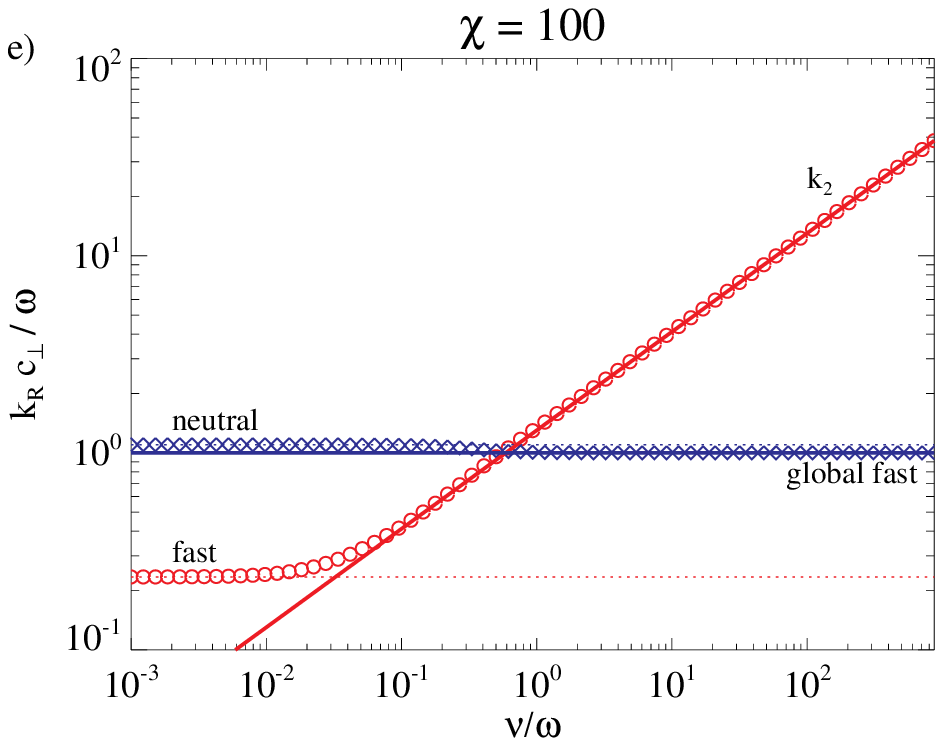}
        \includegraphics[width=0.49\hsize]{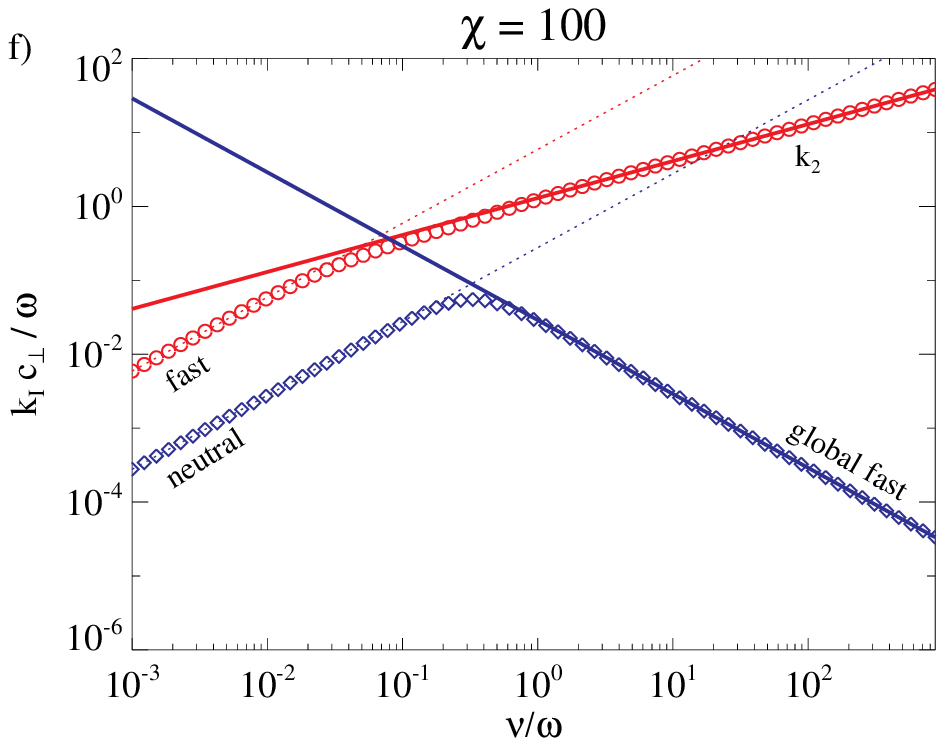} 
        \caption{Normalized wavenumbers (panels a, c, and e) and damping rates (panels b, d, and f) as functions of the coupling degree, $\nu / \omega$, for magneto-acoustic waves propagating in the perpendicular direction to the background magnetic field ($\theta = \pi / 2$), in a hydrogen plasma with $c_{\rm{A}}^{2} = 10 c_{\rm{c}}^{2}$. Top, middle and bottom panels correspond to the density ratios $\chi = 0.01$, $\chi =1$ and $\chi = 100$, respectively. Symbols represent the exact solutions from the dispersion relation, while dotted and solid lines represent the analytical approximations for the weak and strong coupling regimes.}
        \label{fig:perp}
    \end{figure*}
    Now, to check the validity of the approximations previously derived, we compare them with the exact numerical solutions of the dispersion relation. For the sake of simplicity, to perform this comparison we assume that the partially ionized plasma is composed of hydrogen only, so the charged fluid contains electrons and protons and the other fluid contains the neutral hydrogen particles. Assuming charge neutrality of the plasma, the number densities of electrons and protons are identical, so $n_{\rm{e}} = n_{\rm{p}}$, and the number density of the charged fluid is given by $n_{\rm{c}} = n_{\rm{e}} + n_{\rm{p}} = 2 n_{\rm{p}}$. The mass densities of the two fluids in the equilibrium state are then $\rho_{\rm{c0}} = n_{\rm{p}} m_{\rm{p}}$ (since the contribution of the electrons to the density is negligible) and $\rho_{\rm{n0}} = n_{\rm{n}} m_{\rm{p}}$, where $m_{\rm{p}}$ is the proton mass and $n_{\rm{n}}$ is the number density of neutral particles. In addition, we assume that the pressure, number densities and temperature of each fluid are related by the ideal gas law, so 
    \begin{equation}
        P_{\rm{c0}} = n_{\rm{c}} k_{\rm{B}} T_{\rm{c0}} = 2 \frac{\rho_{\rm{c0}}}{m_{\rm{p}}} k_{\rm{B}} T_{\rm{c0}}
    \end{equation}
    and 
    \begin{equation}
        P_{\rm{n0}}= n_{\rm{n}} k_{\rm{B}} T_{\rm{n0}} = \frac{\rho_{\rm{n0}}}{m_{\rm{p}}} k_{\rm{B}} T_{\rm{n0}},
    \end{equation}
    where $k_{\rm{B}}$ is the Boltzmann constant and $T_{\rm{c0}}$ and $T_{\rm{n0}}$ are the equilibrium temperatures of the charged and the neutral fluids, respectively. We also impose that there is a strong thermal coupling between the two fluids, so their equilibrium temperatures are the same. Therefore, the sound speeds of charges and neutrals are related by
    \begin{equation}
        c_{\rm{c}}^{2} = 2 c_{\rm{n}}^{2}.
    \end{equation}
    
    The nature and behavior of the different wave modes depend on the relations between the three characteristic speeds ($c_{\rm{A}}$, $c_{\rm{c}}$, and $c_{\rm{n}}$), on the ratio between the neutral and charged densities ($\chi$), and the coupling degree (determined by the ratio between the wave frequency $\omega$ and the collision frequencies $\nu_{\rm{cn}}$ and $\nu_{\rm{nc}}$). For the present study, it is useful to define the average collision frequency as \citep{Soler2013ApJS..209...16S}
    \begin{equation}
        \nu = \frac{\rho_{\rm{c0}} \nu_{\rm{cn}} + \rho_{\rm{n0}} \nu_{\rm{nc}}}{\rho_{\rm{c0}} + \rho_{\rm{n0}}} = \frac{2 \chi}{1 + \chi}\nu_{\rm{nc}},
    \end{equation}
    and represent the results in terms of this variable.
    
    In Fig. \ref{fig:perp} we show the numerical solutions of the dispersion relation given by Eq. (\ref{eq:ma_dr_perp}) as functions of the coupling degree (represented by the normalized average collision frequency, $\nu / \omega$) for the case with $c_{\rm{A}}^{2} = 10 c_{\rm{c}}^{2}$ and several values of the ionization degree. For the sake of simplicity, only the solutions for forward propagating waves (that is, with $\{k_{\rm{R}}, k_{\rm{I}} \} > 0$) are displayed (the solutions for backward propagating waves have $\{k_{\rm{R}},k_{\rm{I}}\} < 0$ but the same absolute value). In panels $a)$ and $b)$, we show the wavenumber and the damping rates, respectively, for a strongly ionized plasma, with a neutral-to-charges density ratio given by $\chi = 0.01$, so $c_{\rm{A}}^{2} \gg \chi c_{\rm{n}}^{2}$. The wavenumber of the neutral-acoustic mode is larger than that of the fast mode, which is explained by its smaller phase speed. In the limit of weak collisional coupling ($\nu / \omega \ll 1$), the damping rates of both modes increase with the coupling degree. However, their behavior differs as the parameter $\nu / \omega$ increases: the neutral mode becomes the strongly attenuated mode denoted as $k_{\rm{2}}$, whose wavenumber and damping rate grow with the coupling degree; on the other hand, the global fast mode can be directly related to the fast mode of the charged fluid, with a damping rate that decreases as the two fluids become more coupled.

    The case of a partially ionized plasma, with $\chi = 1$, represented in panels $c)$ and $d)$ of Fig. \ref{fig:perp} has a qualitatively similar behavior to the case with $\chi = 0.01$ described in the previous paragraph. The main variations are that now the wavenumber of the global fast mode differs more clearly from that of the charged-fluid fast mode due to a stronger contribution of the inertia of the neutral fluid, and that the upper limit of the weak coupling regime is displaced towards larger values of the coupling degree.

    Then, the results for a weakly ionized plasma, with $\chi = 100$ (and $\chi c_{\rm{n}}^{2} \gg c_{\rm{A}}^{2} + c_{\rm{c}}^{2}$) are displayed in Fig. \ref{fig:perp}$e), f)$. Under these physical conditions, the main contribution to the global fast mode comes from the neutral fluid, meaning that the global fast mode is mainly a neutral-acoustic mode, while the fast mode of the charged fluid is strongly attenuated due to the collisional interaction.

    In addition, Fig. \ref{fig:perp} includes the results of the analytical approximations for the weak coupling limit given by Eqs. (\ref{eq:ksol_perp_weak_fast}) and (\ref{eq:k2_perp_weak_sound}), as dotted lines, and of the approximations for the strong coupling limit given by Eqs. (\ref{eq:perp_strong_krd0}), (\ref{eq:perp_strong_kId1}) and (\ref{eq:sols_perp_strong2}), as solid lines. These approximations show a very good agreement with the exact numerical solutions computed from the full dispersion relation.
    
\subsection{Acoustic waves}
    It has been shown in previous works \citep[see, e.g.,][]{Vranjes2010PhPl...17b2104V,MartinezGomez2018ApJ...856...16M,Popescu2019A&A...627A..25P,Molevich2024PhPl...31d2115M} that the dispersion relation for acoustic waves in partially ionized two-fluid plasmas has the same form as the third branch from Eq. (\ref{eq:ma_dr_perp}) or as Eq. (\ref{eq:ma_perp}) but with $c_{\rm{A}} = 0$. Therefore, the approximate solutions for this kind of waves can be straightforwardly extracted from those presented in Sections \ref{sec:ma_perp_weak} and \ref{sec:ma_perp_coupled}, and the same discussion applies but using the terms ``global acoustic mode'' instead of ``global fast mode'' and ``acoustic mode of the charged fluid'' instead of ``fast mode of the charged fluid''. In addition, if we set $c_{\rm{A}} = 0$ in Eq. (\ref{eq:perp_strong_wId1}), it can be shown that this expression for the damping rates of acoustic waves generated by an impulsive driver agrees with those provided in Table 2 of Ref. \onlinecite{Molevich2024PhPl...31d2115M} (although a different notation has been used).
    
\subsection{Magneto-acoustic waves: oblique propagation} \label{sec:ma_oblique}
    In this section, we investigate the solutions of the dispersion relation for magneto-acoustic waves propagating at an arbitrary angle $\theta \neq \pm \pi/2$ with respect to the background magnetic field. Since the case of waves generated by an impulsive driver (with a real $k$ but a complex $\omega$) has already been described in detail in Ref. \onlinecite{Soler2024RSPTA.38230223S}, here we only examine the case with a real wave frequency but a complex wavenumber. Therefore, the dispersion relation given by Eq. (\ref{eq:ma_dr}) or Eq. (\ref{eq:ma_dr_full}) will provide six different solutions for $k$.
    
\subsubsection{Weak coupling regime for oblique propagation} \label{sec:ma_weakly_oblique}
    Once more, we start by studying the limit of weak collisional coupling. Assuming that $\{\nu_{\rm{cn}}, \nu_{\rm{nc}} \} \ll \omega$, the last term from Eq. (\ref{eq:ma_dr}) can be dropped and the dispersion relation is now given by $\mathcal{D}\left(\omega, \bm{k} \right) \approx D_{\rm{c}}\left(\omega, \bm{k} \right) D_{\rm{n}}\left(\omega, \bm{k} \right) = 0$. Again, we can solve two independent dispersion relations, one related to each fluid. The equation $D_{\rm{n}}\left(\omega, \bm{k} \right) = 0$ provides the same two solutions associated with the neutral fluid already described in Section \ref{sec:ma_perp_weak}, so there is no need to repeat those results here. Thus, we now focus on the four solutions coming from the equation $D_{\rm{c}}\left(\omega, \bm{k} \right) = 0$, which can be rewritten as
    \begin{gather} \label{eq:ma_oblique_weak}
        D_{\rm{c}}\left(\omega, \bm{k}\right) = \Big[\omega^{3} \left(\omega + i \chi \nu_{\rm{nc}} \right) - \omega^{2} k^{2}\left(c_{\rm{A}}^{2} + c_{\rm{c}}^{2} \right) \Big] \nonumber \\
        \times \Big(\omega + i \left(1 + \chi \right) \nu_{\rm{nc}} \Big) + \left(\omega + i \nu_{\rm{nc}} \right)k^{4}c_{\rm{A}}^{2}c_{\rm{c}}^{2} \cos^{2} \theta = 0,
    \end{gather}
    where again the relation $\nu_{\rm{cn}} = \chi \nu_{\rm{nc}}$ has been applied.

    Then, we split this new expression as $D_{\rm{c}} \left(\omega, \bm{k} \right) =  D_{\rm{R}}^{\rm{c}} + i D_{\rm{I}}^{\rm{c}}$, with
    \begin{eqnarray}
        D_{\rm{R}}^{\rm{c}} &=& \omega \left[\omega^{4} - k^{2} \left(c_{\rm{A}}^{2} + c_{\rm{c}}^{2} \right) \omega^{2} + k^{4} c_{\rm{A}}^{2} c_{\rm{c}}^{2} \cos^{2} \theta \right] \nonumber \\
        &-& \left(\chi + \chi^{2} \right)\nu_{\rm{nc}}^{2} \omega^{3},
        \label{eq:ma_dRc_weak}
    \end{eqnarray}
    and
    \begin{eqnarray}
        D_{\rm{I}}^{\rm{c}} &=& \nu_{\rm{nc}} \Big[\left(1 + 2 \chi \right) \omega^{4} - \left(1 + \chi \right) k^{2} \left(c_{\rm{A}}^{2} + c_{\rm{c}}^{2} \right) \omega^{2} \nonumber \\
        &+&k^{4} c_{\rm{A}}^{2} c_{\rm{c}}^{2} \cos^{2} \theta \Big], \label{eq:ma_dIc_weak}
    \end{eqnarray}
    in order to apply the method from Ref. \onlinecite{Tagger1995A&A...299..940T}, in the same way as it has been done in Section \ref{sec:ma_perp_coupled}. Since we have that $\chi \nu_{\rm{nc}} \ll \omega$, the last term from Eq. (\ref{eq:ma_dRc_weak}) can be neglected. If we solve the equation $D_{\rm{R}}^{\rm{c}} = 0$ for a real value of $\omega$ we then get the following expression for the real part of the wavenumber:
    \begin{equation} \label{eq:ma_oblique_weak_kr}
        k_{\rm{R}} = \pm \frac{\omega \sec \theta}{\sqrt{2} c_{\rm{A}}c_{\rm{c}}}\Bigg[\left(c_{\rm{A}}^{2} + c_{\rm{c}}^{2}\right)\Bigg(1 \pm \sqrt{1 - \frac{4 c_{\rm{A}}^{2}c_{\rm{c}}^{2} \cos^{2} \theta}{\left(c_{\rm{A}}^{2} + c_{\rm{c}}^{2} \right)^{2}}}\Bigg) \Bigg]^{1/2},
    \end{equation}
    where the outer ``$\pm$'' signs distinguish between the forward and backward propagating modes, and the inner ``$\pm$'' signs distinguish between the slow and fast modes. This is the classical result for slow and fast magneto-acoustic waves in a fully ionized plasma \citep[see, e.g.,][]{Goedbloed2004prma.book.....G}.

    In addition, Eq. (\ref{eq:ma_dIc_weak}) simplifies to
    \begin{equation}
        D_{\rm{I}}^{\rm{c}} = - \chi \nu_{\rm{nc}} \omega^{2} \left[ k^{2} \left(c_{\rm{A}}^{2} + c_{\rm{c}}^{2} \right) - 2 \omega^{2} \right]
    \end{equation}
    and the damping rate is given by
    \begin{equation} \label{eq:ma_oblique_weak_kI}
        k_{\rm{I}} \approx - \frac{D_{\rm{I}}^{\rm{c}}}{\partial D_{\rm{R}}^{\rm{c}} / \partial k}\Bigg|_{\omega,k_{\rm{R}}} = \frac{\nu_{\rm{cn}}}{2}\frac{k_{\rm{R}}}{\omega}.        
    \end{equation}

    Equations (\ref{eq:ma_oblique_weak_kr}) and (\ref{eq:ma_oblique_weak_kI}) are valid for any value of the characteristic speeds different from zero. However, even simpler expressions can be found in certain interesting limits. For instance, in the limit $c_{\rm{A}}^{2} \gg c_{\rm{c}}^2$, from Eq. (\ref{eq:ma_oblique_weak_kr}) we get that
    \begin{equation} \label{eq:ma_oblique_weak_lowbeta_kr}
        k_{\rm{R}} \approx \pm \frac{\omega}{c_{\rm{A}}} \quad \text{and} \quad k_{\rm{R}} \approx \pm \frac{\omega \sec \theta}{c_{\rm{c}}},
    \end{equation}
    for the fast and slow modes, respectively. On the other hand, in the limit $c_{\rm{A}}^{2} \ll c_{\rm{c}}^{2}$ the real part of the wavenumber of the fast and the slow modes are given by
    \begin{equation} \label{eq:ma_oblique_weak_highbeta_kr}
        k_{\rm{R}} \approx \pm \frac{\omega}{c_{\rm{c}}} \quad \text{and} \quad k_{\rm{R}} \approx \pm \frac{\omega \sec \theta}{c_{\rm{A}}},
    \end{equation}
    respectively. The corresponding damping rates in both limits can be computed by combining Eqs. (\ref{eq:ma_oblique_weak_lowbeta_kr}) and (\ref{eq:ma_oblique_weak_highbeta_kr}) with Eq. (\ref{eq:ma_oblique_weak_kI}).

    The expressions obtained from Eq. (\ref{eq:ma_oblique_weak_lowbeta_kr}) and its combination with Eq. (\ref{eq:ma_oblique_weak_kI}) are consistent with the ones that can be derived from Equation (4) from Ref. \onlinecite{CallyGomezMiguez2023ApJ...946..108C} in the limit $\nu_{\rm{cn}} \ll \omega$.

\subsubsection{Strong coupling regime for oblique propagation} \label{sec:ma_strong_oblique}
    The first step to derive approximate expressions for the limit of strong collisional coupling is to expand either Eq. (\ref{eq:ma_dr}) or Eq. (\ref{eq:ma_dr_full}) in powers of the collision frequencies and to only retain the terms that are proportional to $\nu_{\rm{cn}}$ or $\nu_{\rm{nc}}$, while discarding the remaining ones. Then, the resulting equation can be split into its real and imaginary parts in order to apply the method from Ref. \onlinecite{Tagger1995A&A...299..940T}. An alternative method is to write the wavenumber as $k = k_{\rm{R}} + i k_{\rm{I}}$ and assume that $k_{\rm{I}} \ll k_{\rm{R}}$, so the second and higher order powers of $k_{\rm{I}}$ can be neglected. Either way, we will reach to the following equation for the real part of the wavenumber:
    \begin{gather} 
        k_{\rm{R}}^{4} - \frac{\left(1 + \chi \right) \left(c_{\rm{A}}^{2} + c_{\rm{c}}^{2} + \chi c_{\rm{n}}^{2} \right) \omega^{2}}{c_{\rm{A}}^{2} \left(c_{\rm{c}}^{2} + \chi c_{\rm{n}}^{2} \right) \cos^{2} \theta} k_{\rm{R}}^{2} \nonumber \\
        + \frac{\left(1 + \chi \right)^{2}}{c_{\rm{A}}^{2} \left(c_{\rm{c}}^{2} + \chi c_{\rm{n}}^{2} \right) \cos^{2} \theta} \omega^{4} =0,
        \label{eq:kr4}
    \end{gather}
    or, equivalently,
    \begin{equation} \label{eq:kr4_global}
        k_{\rm{R}}^{4} - \frac{\omega^{2} \left(a^{2} + c^{2} \right)}{a^{2} c^{2} \cos^{2} \theta} k_{\rm{R}}^{2} + \frac{\omega^{4}}{a^{2}c^{2} \cos^{2} \theta} = 0,
    \end{equation}
    whose solution is given by    
    \begin{equation} \label{eq:ma_oblique_strong_kr}
        k_{\rm{R}} = \pm \frac{\omega \sec \theta}{\sqrt{2} a c}\Bigg[\left(a^{2} + c^{2}\right)\Bigg(1 \pm \sqrt{1 - \frac{4 a^{2}c^{2} \cos^{2} \theta}{\left(a^{2} + c^{2} \right)^{2}}}\Bigg) \Bigg]^{1/2}.
    \end{equation}

    Equation (\ref{eq:ma_oblique_strong_kr}) has the same functional form as Eq. (\ref{eq:ma_oblique_weak_kr}), but substituting $c_{\rm{A}}$ and $c_{\rm{c}}$ by $a$ and $c$, respectively, and it provides the real part of the wavenumber of the global fast and slow magneto-acoustic waves, taking into account the contribution from both fluids.

    The corresponding damping rates are computed as
    \begin{equation} \label{eq:ma_oblique_strong_ki}
        k_{\rm{I}} \approx \frac{\mathcal{A}_{\rm{kI}} \left(\omega, k_{\rm{R}} \right)}{\mathcal{B}_{\rm{kI}} \left(\omega, k_{\rm{R}} \right)},
    \end{equation}
    where
    \begin{gather}
        \mathcal{A}_{\rm{kI}} \left(\omega, k_{\rm{R}} \right)= \left(c_{\rm{A}}^{2} + c_{\rm{c}}^{2} \right) c_{\rm{n}}^{2} k_{\rm{R}}^{4} \left(1 + \chi \right) \omega^{2} \nonumber \\
        - k_{\rm{R}}^{2} \Big(c_{\rm{A}}^{2} \left(2 + \chi \right) + c_{\rm{c}}^{2} \left(2 + \chi \right) + c_{\rm{n}}^{2} \left(1 + 2 \chi \right) \Big) \omega^{4} \nonumber \\
        + 2 \left(1 + \chi \right) \omega^{6} + c_{\rm{A}}^{2} c_{\rm{c}}^{2} k_{\rm{R}}^{4} \left(2 \omega^{2} - c_{\rm{n}}^{2} k_{\rm{R}}^{2} \right) \cos^{2} \theta,
    \end{gather}
    \begin{gather}
        \mathcal{B}_{\rm{kI}} \left(\omega, k_{\rm{R}} \right)= \nu_{\rm{nc}} \omega k_{\rm{R}} \left(1 + \chi \right)^{2} \nonumber \\
        \times \left[4 a^2 c^2 k_{\rm{R}}^{2} \cos^{2} \theta - 2 \omega^{2} \left(a^{2} + c^{2} \right) \right],
    \end{gather}
    and $k_{\rm{R}}$ is given by Eq. (\ref{eq:ma_oblique_strong_kr}). In general, the resulting expression is very convoluted and it does not seem to provide an advantage with respect to numerically solving the full dispersion relation. However, it can be used to derive simpler approximations for some particular cases of usual interest in the research of partially ionized plasmas.
    
    For instance, in the limit of strongly ionized plasmas with $c_{\rm{A}}^{2} \gg c_{\rm{c}}^{2} \gg \chi c_{\rm{n}}^{2}$, Eqs. (\ref{eq:ma_oblique_strong_kr}) and (\ref{eq:ma_oblique_strong_ki}) reduce to
    \begin{equation} \label{eq:kr_fast_aa}
        k_{\rm{R,f}} \approx \pm \frac{\omega}{a} = \pm \frac{\sqrt{1 + \chi}}{c_{\rm{A}}} \omega,
    \end{equation}
    \begin{equation} \label{eq:kr_slow_ceff}
        k_{\rm{R,s}} \approx \pm \frac{\omega}{c \cos \theta} = \pm \sqrt{\frac{1 + \chi}{c_{\rm{c}}^{2} + \chi c_{\rm{n}}^{2}}} \frac{\omega}{\cos \theta},
    \end{equation}
    \begin{equation} \label{eq:ki_fast_aa}
        k_{\rm{I,f}} \approx \pm \frac{\chi \omega^{2}}{2 \nu_{\rm{nc}} \sqrt{1 + \chi}} \frac{1}{c_{\rm{A}}},
    \end{equation}
    and
    \begin{equation} \label{eq:ki_slow_ceff}
        k_{\rm{I,s}} \approx \pm \frac{\chi \omega^{2} \Big(c_{\rm{c}}^{4} - 2 c_{\rm{c}}^{2} c_{\rm{n}}^{2} - \chi c_{\rm{n}}^{4} (2 + \chi) + c_{\rm{n}}^{4} (1 + \chi)^{2} \sec^{2} \theta \Big)}{2 \nu_{\rm{nc}} \sqrt{1 + \chi} \left(c_{\rm{c}}^{2} + \chi c_{\rm{n}}^{2} \right)^{5/2} \cos \theta},
    \end{equation}
    where the subscripts $f$ and $s$ refer to the fast and slow modes respectively.

    In contrast, if the gas pressure of the charged fluid dominates the dynamics of the plasma, with $c_{\rm{c}}^{2} \gg c_{\rm{A}}^{2} \gg \chi c_{\rm{n}}^{2}$, we find the following approximations:
    \begin{equation} \label{eq:kr_fast_largecc}
        k_{\rm{R,f}} \approx \pm \frac{\omega}{c} \approx \pm \frac{\omega}{c_{\rm{c}}},
    \end{equation}
    \begin{equation} \label{eq:kr_slow_largecc}
        k_{\rm{R,s}} \approx \pm \Big(\frac{\sqrt{1 + \chi}}{c_{\rm{A}}\cos \theta} + \frac{c_{\rm{A}}\sqrt{1 + \chi}}{2c_{\rm{c}}^{2}} \frac{\sin^{2} \theta}{\cos \theta} \Big) \omega,
    \end{equation}
    \begin{equation} \label{eq:ki_fast_largecc}
        k_{\rm{I,f}} \approx \pm \frac{\chi \omega^{2}}{2 \nu_{\rm{nc}} \sqrt{1 + \chi}} \frac{\left(c_{\rm{c}}^{2} - c_{\rm{n}}^{2} \right)^{2}}{\left(c_{\rm{c}}^{2} + \chi c_{\rm{n}}^{2} \right)^{5/2}},
    \end{equation}
    and
    \begin{equation} \label{eq:ki_slow_largecc}
        k_{\rm{I,s}} \approx \pm \frac{\chi \omega^{2}}{2 \nu_{\rm{nc}} \sqrt{1 + \chi}} \frac{1}{c_{\rm{A}} \cos \theta}.
    \end{equation}
    
    In the case of a weakly ionized plasma with $\chi c_{\rm{n}}^{2} \gg c_{\rm{c}}^{2} \gg c_{\rm{A}}^{2}$, the wavenumbers and damping rates of the global fast and slow modes are given by
    \begin{equation} \label{eq:kr_fast_ceff}
        k_{\rm{R,f}} \approx \pm \frac{\omega}{c} = \pm \sqrt{\frac{1 + \chi}{c_{\rm{c}}^{2} + \chi c_{\rm{n}}^{2}}} \omega \approx \pm \frac{\omega}{c_{\rm{n}}},
    \end{equation}
    \begin{equation} \label{eq:kr_slow_aa}
        k_{\rm{R,s}} \approx \pm \frac{\omega}{a \cos \theta} = \pm \frac{\sqrt{1 + \chi}}{c_{\rm{A}} \cos \theta} \omega,
    \end{equation}
    \begin{equation} \label{eq:ki_slow_aa}
        k_{\rm{I,s}} \approx \pm \frac{\omega^{2}\Big(\chi c_{\rm{c}}^{2} - c_{\rm{n}}^{2} \left(1 + 2 \chi \right) + c_{\rm{n}}^{2} \left(1 + \chi \right)^{2} \sec^{2} \theta \Big)}{2 \nu_{\rm{nc}} \sqrt{1 + \chi} \left(c_{\rm{c}}^{2} + \chi c_{\rm{n}}^{2} \right) c_{\rm{A}} \cos \theta}.
    \end{equation}
    and the damping rate of the global fast mode ($k_{\rm{I,f}}$) is again given by Eq. (\ref{eq:ki_fast_largecc}).

    Furthermore, if we set $\theta = 0$ in the previous formulas, one of the damping rates for waves propagating along the parallel direction to the background magnetic field has the same expression as Eq. (\ref{eq:ki_fast_aa}), which coincides with the damping rate for propagating Alfvén waves in two-fluid partially ionized plasma, as shown in Refs. \onlinecite{Soler2013ApJ...767..171S,MartinezGomez2025ApJ...982....4M}. The remaining damping rate is identical to Eq. (\ref{eq:ki_fast_largecc}), which is the same expression that would be obtained for acoustic waves by setting $c_{\rm{A}} = 0$ in Eq. (\ref{eq:perp_strong_kId1}).

    The approximations discussed in the previous paragraphs correspond to four of the six solutions of the dispersion relation. To find the approximate expressions for the remaining two solutions, we follow the method used in Refs. \onlinecite{CallyGomezMiguez2023ApJ...946..108C} and we first rewrite the general dispersion relation given by Eq. (\ref{eq:ma_dr_full}) as a bi-cubic equation,
    \begin{equation} \label{eq:bicubic}
        b_{\rm{3}} r^{3} + b_{\rm{2}} r^{2} + b_{\rm{1}} r + b_{\rm{0}} = 0,
    \end{equation}
    where $r = k^{2}$. If $r_{\rm{1}} \equiv k_{\rm{1}}^{2}$, $r_{\rm{2}} \equiv k_{\rm{2}}^{2}$, and $r_{\rm{3}} \equiv k_{\rm{3}}^{2}$ are the roots of Eq. (\ref{eq:bicubic}), they fulfill the following Vieta's formula:
    \begin{equation} \label{eq:r1r2r3}
        r_{\rm{1}} r_{\rm{2}} r_{\rm{3}} = k_{\rm{1}}^{2} k_{\rm{2}}^{2} k_{\rm{3}}^{3} = -\frac{b_{\rm{0}}}{b_{\rm{3}}}.
    \end{equation}
    Comparing Eqs. (\ref{eq:ma_dr_full}) and (\ref{eq:bicubic}) we find that the coefficients $b_{\rm{0}}$ and $b_{\rm{3}}$ are given by
    \begin{equation}
        b_{\rm{0}} = \omega^{5} \Big( \omega + i \left(1 + \chi \right) \nu_{\rm{nc}} \Big)^{2}
    \end{equation}
    and
    \begin{equation}
        b_{\rm{3}} = -\left(\omega + i \nu_{\rm{nc}} \right) c_{\rm{A}}^{2} c_{\rm{c}}^{2} c_{\rm{n}}^{2} \cos^{2} \theta.
    \end{equation}
    Imposing the condition of strong collisional coupling ($\nu_{\rm{nc}} \gg \omega$), from Eq. (\ref{eq:r1r2r3}) we arrive to the following relation for the roots of Eq. (\ref{eq:bicubic}):
    \begin{equation} \label{eq:k1k2k3}
        k_{\rm{1}}^{2} k_{\rm{2}}^{2} k_{\rm{3}}^{3} \approx \frac{i \omega^{5} \nu_{\rm{nc}} \left(1 + \chi \right)^{2}}{c_{\rm{A}}^{2} c_{\rm{c}}^{2} c_{\rm{n}}^{2} \cos^{2} \theta}.
    \end{equation}
    Then, we assume that $k_{\rm{1}}^{2}$ and $k_{\rm{2}}^{2}$ correspond to the global fast and slow modes in the strong coupling regime, which will be solutions of Eq. (\ref{eq:kr4}) and, therefore, will fulfill the Vieta's relation
     \begin{equation} \label{eq:k1k2_general}
        k_{1}^{2}k_{2}^{2} = \frac{\left(1 + \chi \right)^{2}}{c_{\rm{A}}^{2} \left(c_{\rm{c}}^{2} + \chi c_{\rm{n}}^{2} \right) \cos^{2} \theta} \omega^{4}.   
    \end{equation}
    Finally, by substituting Eq. (\ref{eq:k1k2_general}) into Eq. (\ref{eq:k1k2k3}), we find that the two remaining solutions are given by
    \begin{equation} \label{eq:k3_general}
        k_{\rm{3}}^{2} \approx \frac{i \nu_{\rm{nc}} \omega \left(c_{\rm{c}}^{2} + \chi c_{\rm{n}}^{2} \right)}{c_{\rm{c}}^{2} c_{\rm{n}}^{2}},
    \end{equation}
    which has the same form as Eq. (\ref{eq:sols_perp_strong2}) but with $c_{\rm{A}}=0$. It can be checked that for the case of a hydrogen only plasma (in which the relations $c_{\rm{c}}^{2} = 2 c_{\rm{n}}^{2}$ and $c_{\rm{n}}^{2} = c^{2} \left(1 + \chi \right) / \left(2 + \chi \right) $ are fulfilled), we recover the expression derived in Ref. \onlinecite{CallyGomezMiguez2023ApJ...946..108C}, namely    
    \begin{equation} \label{eq:k3_hydrogen}
        k_{3}^{2} \approx i \frac{\omega \nu_{\rm{nc
        }} \left(\chi + 2\right)^{2}}{2 c^{2} \left(\chi + 1 \right)}.
    \end{equation}

    Ref. \onlinecite{CallyGomezMiguez2023ApJ...946..108C} stated that the solutions given by Eq. (\ref{eq:k3_hydrogen}) correspond to the neutral-acoustic mode. However, we can see from the more general Eq. (\ref{eq:k3_general}) that the $k_{\rm{3}}$ modes depend on the sound speeds of both fluids, so they cannot be clearly associated to only one of the components of the plasma. In addition, as discussed in Ref. \onlinecite{Soler2013ApJS..209...16S}, the associations between the global or modified magneto-acoustic modes and the individual modes of each separate fluid depend on the ionization degree of the plasma and the relations between the individual characteristic speeds. For instance, in the limit of strong ionization, with $\chi \to 0$, we can find from Eq. (\ref{eq:kr4}) that the global fast and slow modes will fulfill
    \begin{equation} \label{eq:k1k2_chi0}
        k_{\rm{1}}^{2} k_{\rm{2}}^{2} \approx \frac{\omega^{4}}{c_{\rm{A}}^{2}c_{\rm{c}}^{2} \cos^{2} \theta},
    \end{equation}
    so they mainly depend on the parameters of the charged fluid, while Eq. (\ref{eq:k3_general}) reduces to
    \begin{equation} \label{eq:k3_chi0}
        k_{3}^{2} \approx \frac{i \omega \nu_{\rm{nc}}2}{c^{2}} = \frac{i \omega \nu_{\rm{nc}}}{c_{\rm{n}}^{2}},
    \end{equation}
   which can now be associated with the neutral acoustic mode.

   However, in the limit $\chi c_{\rm{n}}^{2} \gg c_{\rm{c}}^{2}$, corresponding to weak ionization of the plasma, the approximate relation for the global fast and slow waves is
   \begin{equation} \label{eq:k1k2_chiinf}
       k_{\rm{1}}^{2} k_{\rm{2}}^{2} \approx \frac{\chi \omega^{4}}{c_{\rm{A}}^{2}c_{\rm{n}}^{2} \cos^{2} \theta},
   \end{equation}
   so these modes strongly depend on the density and the sound speed of the neutral fluid but also on the Alfvén speed of the charged fluid, and Eq. (\ref{eq:k3_general}) becomes
   \begin{equation} \label{eq:k3_chiinf}
        k_{3}^{2} \approx \frac{i \omega \nu_{\rm{nc}}\chi}{c_{\rm{c}}^{2}} = \frac{i \omega \nu_{\rm{cn}}}{c_{\rm{c}}^{2}},
    \end{equation}
    which depends on the parameters of the charged fluid but not on those from the neutral one.

    In a general scenario of strong collisional coupling, the neutral acoustic mode mixes with the fast and slow modes from the charged fluid to result in both the global magneto-acoustic waves \citep[see, e.g.,][]{Soler2013ApJS..209...16S} and the strongly attenuated modes given by Eq. (\ref{eq:k3_general}).

\subsubsection{Parametric study and comparison between analytical approximations and exact numerical solutions of the dispersion relation}
    \begin{figure*}
        \centering
        \includegraphics[width=0.49\hsize]{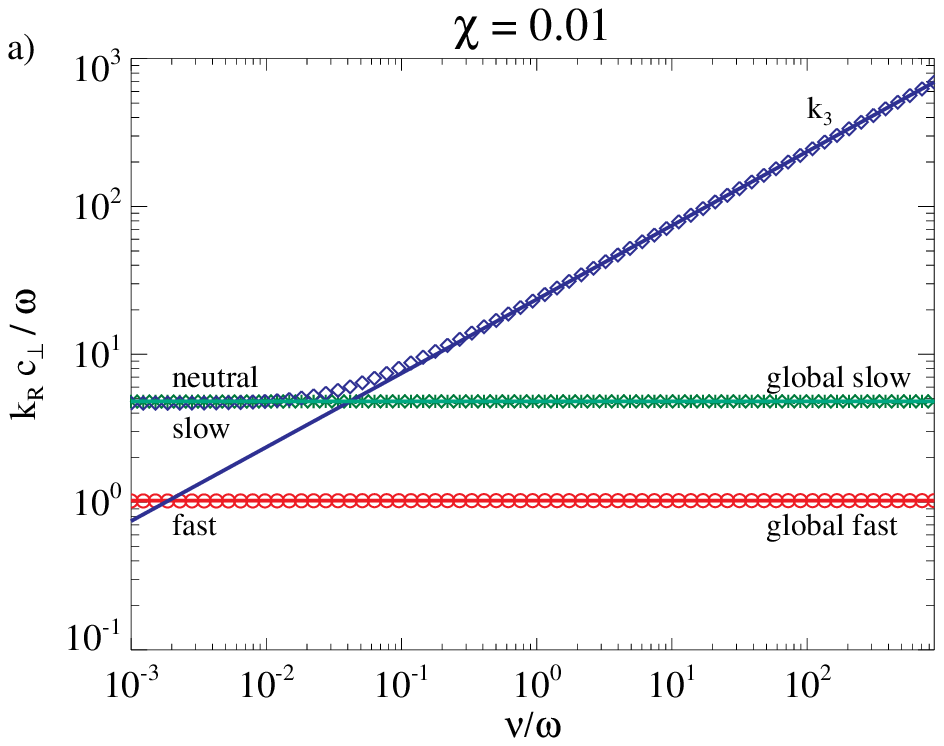}
        \includegraphics[width=0.49\hsize]{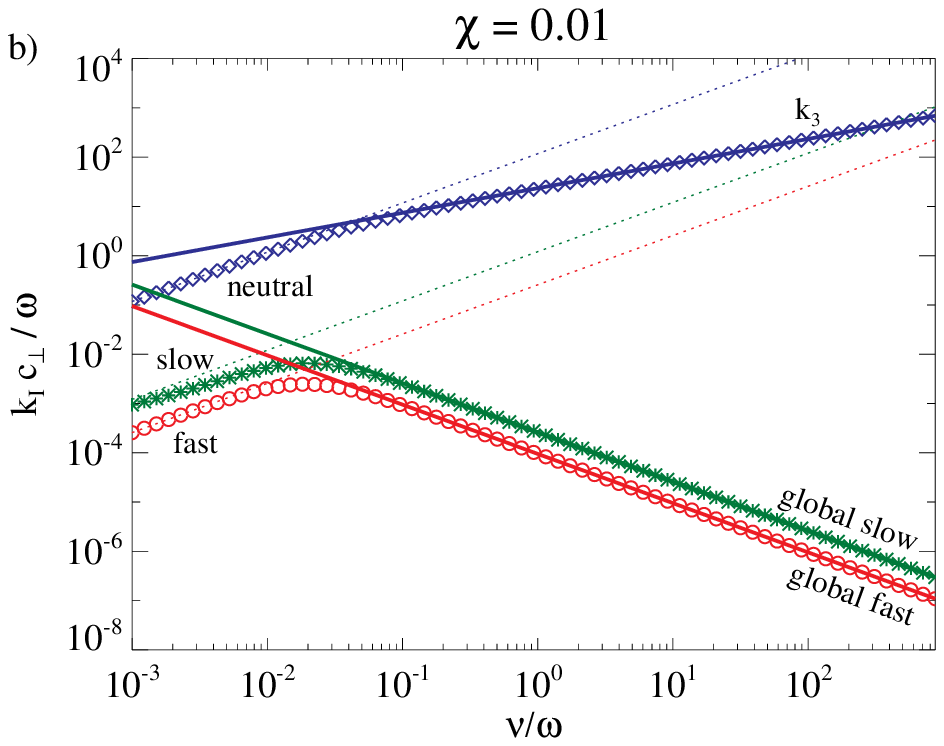} \\
        \includegraphics[width=0.49\hsize]{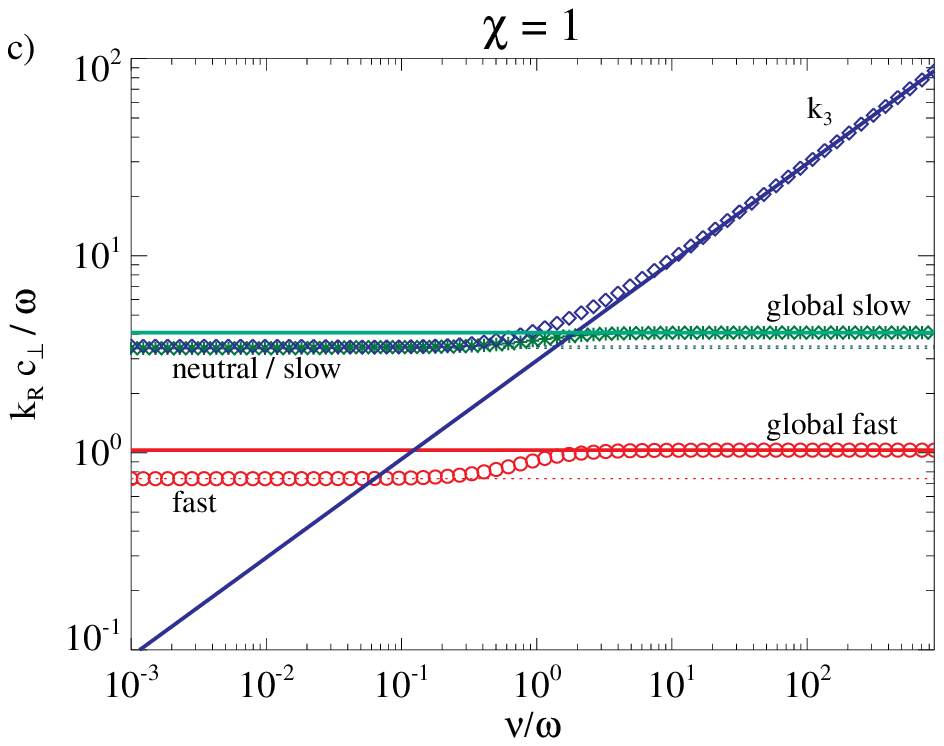}
        \includegraphics[width=0.49\hsize]{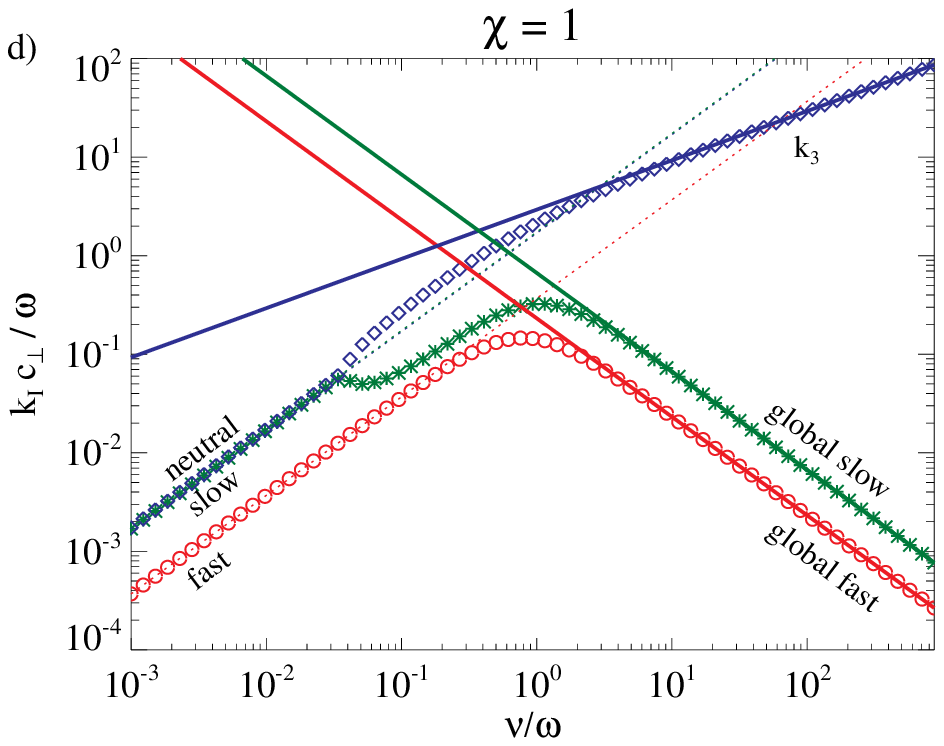} \\
        \includegraphics[width=0.49\hsize]{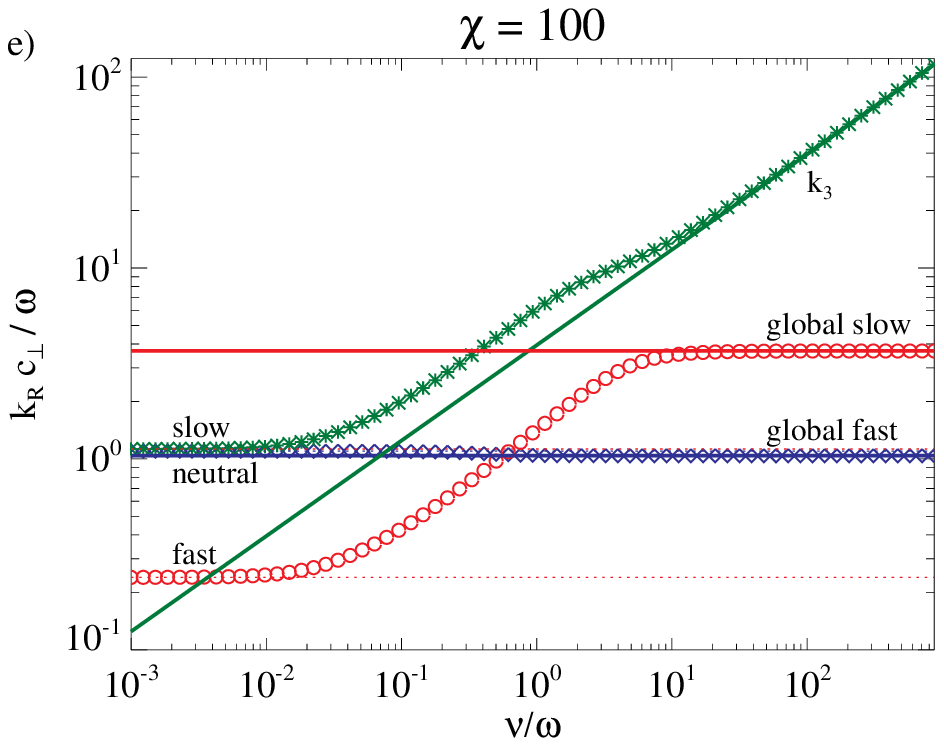}
        \includegraphics[width=0.49\hsize]{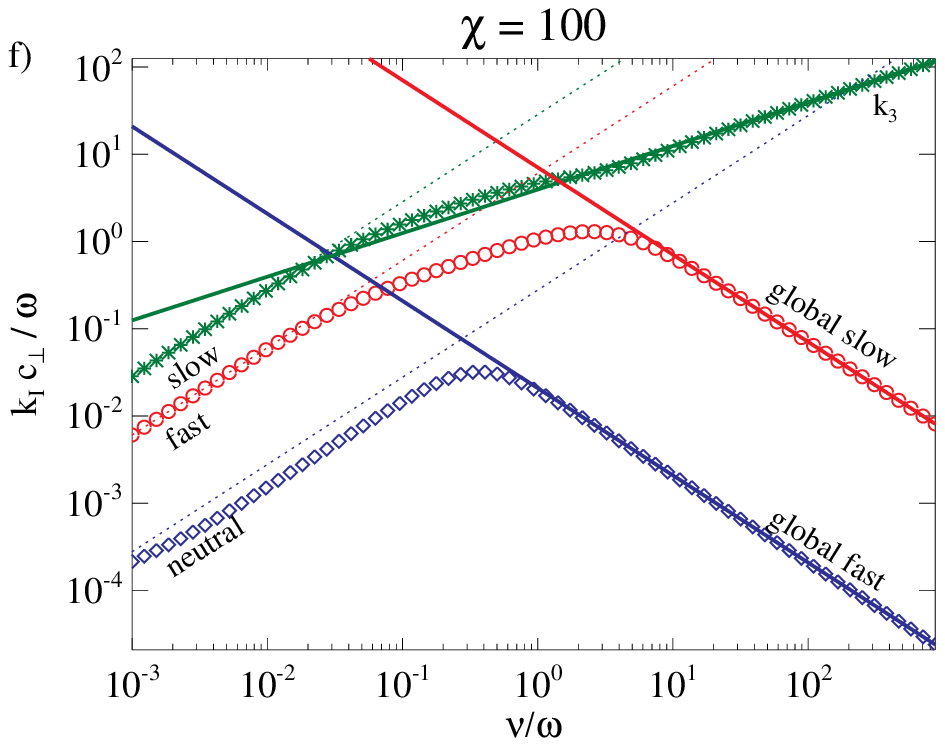} 
        \caption{Normalized wavenumbers (panels a, c, and e) and damping rates (panels b, d, and f) as functions of the coupling degree, $\nu / \omega$, for magneto-acoustic waves propagating at an angle $\theta = \pi /4$ with respect to the background magnetic field, in a hydrogen plasma with $c_{\rm{A}}^{2} = 10 c_{\rm{c}}^{2}$. Top, middle and bottom panels correspond to the density ratios $\chi = 0.01$, $\chi =1$ and $\chi = 100$, respectively. Symbols represent the exact solutions from the dispersion relation, while dotted and solid lines represent the analytical approximations for the weak and strong coupling regimes.}
        \label{fig:oblique}
    \end{figure*}

    \begin{figure*}
        \centering
        \includegraphics[width=0.49\hsize]{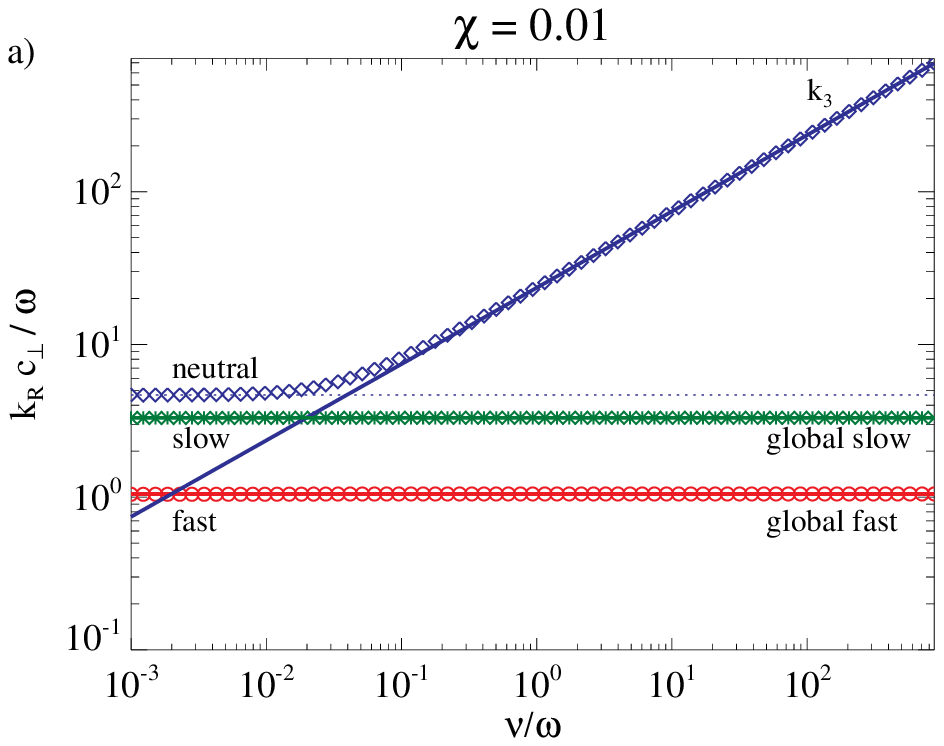}
        \includegraphics[width=0.49\hsize]{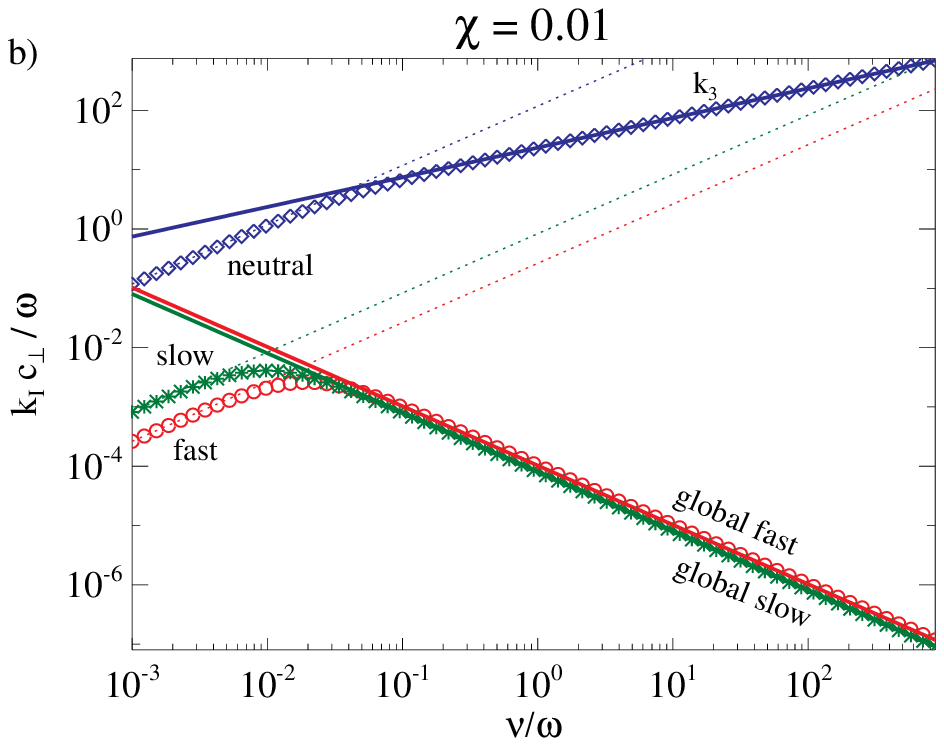} \\
        \includegraphics[width=0.49\hsize]{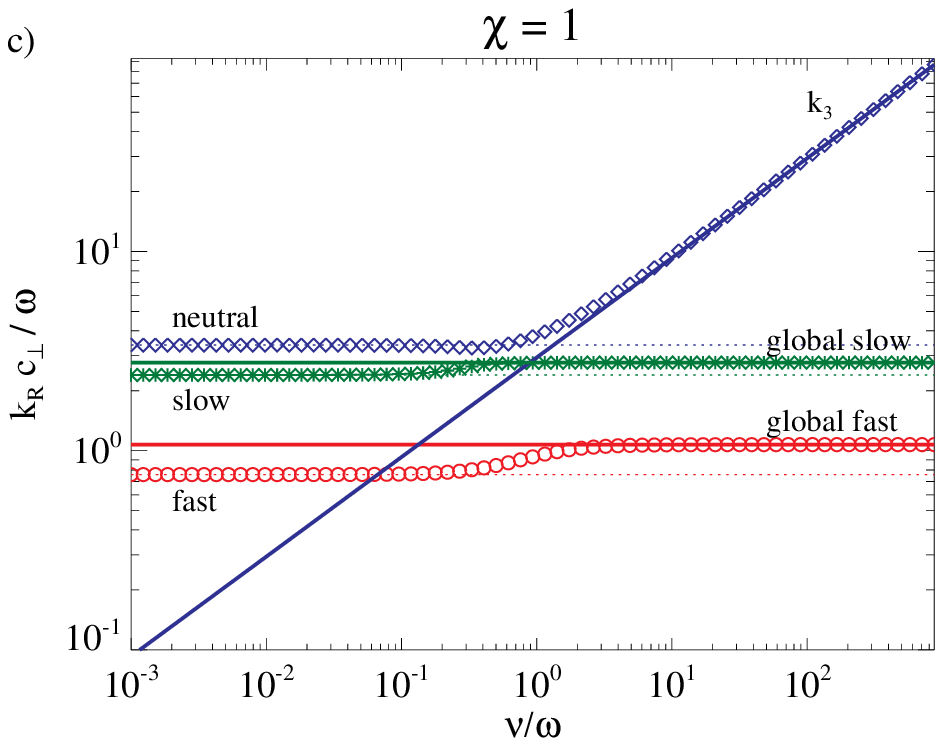}
        \includegraphics[width=0.49\hsize]{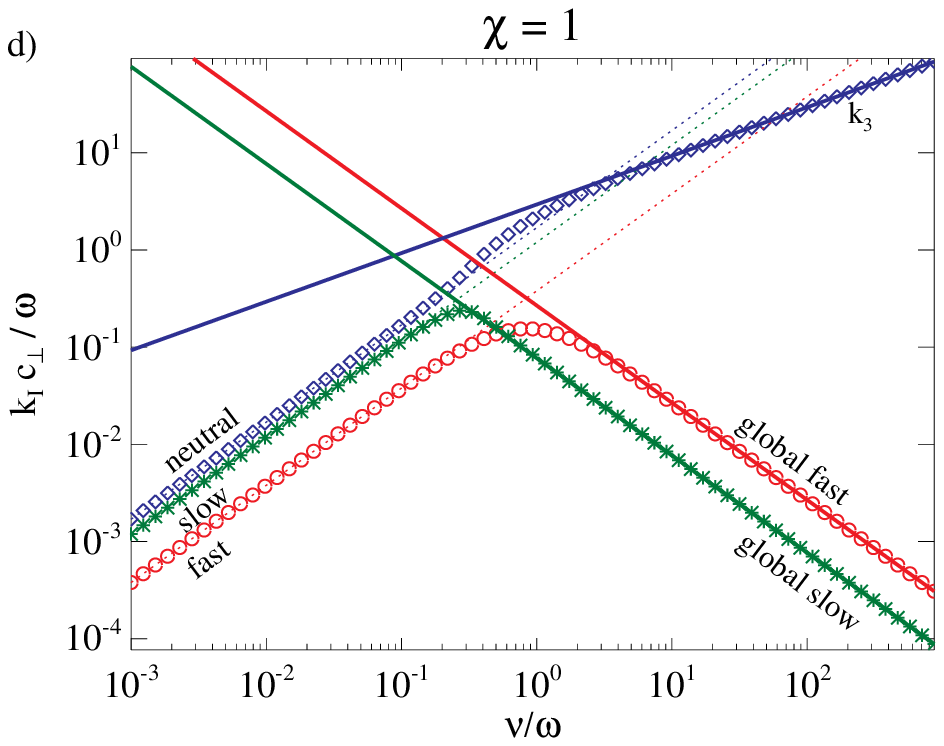} \\
        \includegraphics[width=0.49\hsize]{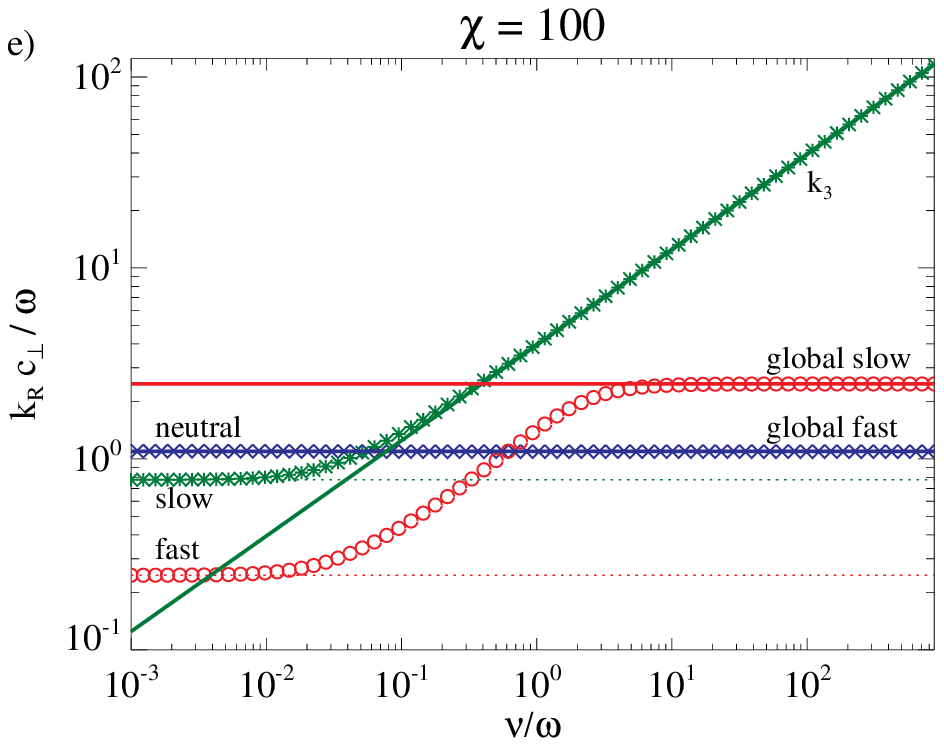}
        \includegraphics[width=0.49\hsize]{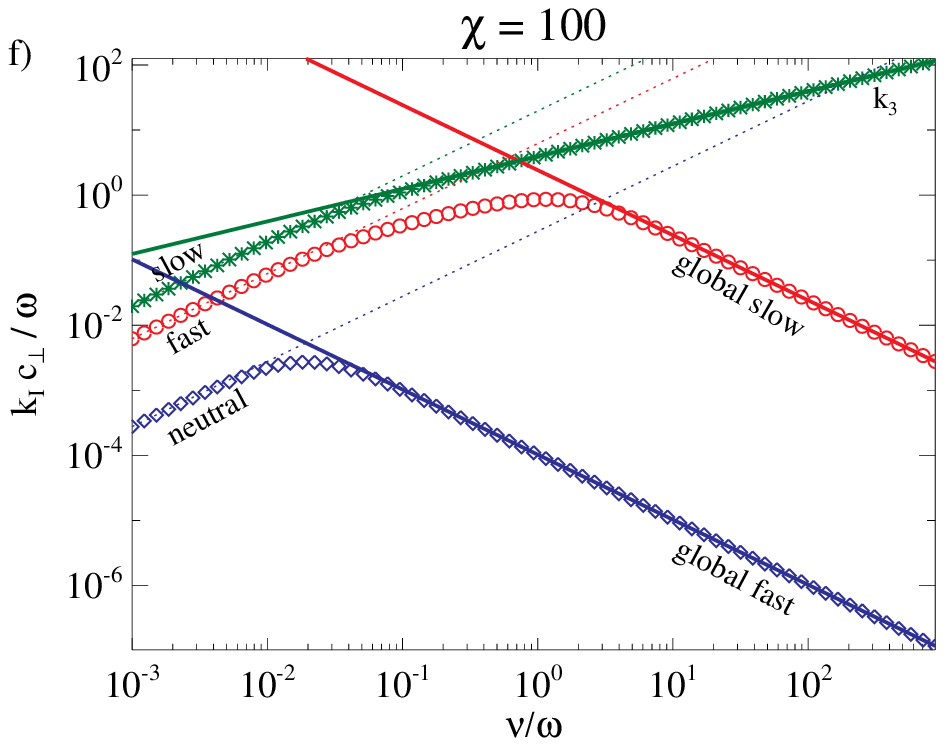} 
        \caption{Normalized wavenumbers (a, c, and e) and damping rates (b, d, and f) as functions of the coupling degree, $\nu / \omega$, for magneto-acoustic waves propagating along the parallel direction to the background magnetic field ($\theta = 0$), in a hydrogen plasma with $c_{\rm{A}}^{2} = 10 c_{\rm{c}}^{2}$. Top, middle and bottom panels correspond to the density ratios $\chi = 0.01$, $\chi =1$ and $\chi = 100$, respectively. Symbols represent the exact solutions from the dispersion relation, while dotted and solid lines represent the analytical approximations for the weak and strong coupling regimes.}
        \label{fig:parallel}
    \end{figure*}
    
    Figure \ref{fig:oblique} shows the dependence on the coupling degree of magneto-acoustic waves propagating at an angle $\theta = \pi / 4$ with respect to the background magnetic field. Again, a hydrogen plasma with $c_{\rm{A}}^{2} = 10 c_{\rm{c}}^{2}$ is considered. Symbols represent the exact numerical solutions from Eq. (\ref{eq:ma_dr_full}). Dotted lines correspond to the weak coupling approximations given by Eqs. (\ref{eq:ksol_perp_weak_sound}), (\ref{eq:ma_oblique_weak_kr}) and (\ref{eq:ma_oblique_weak_kI}). Finally, solid lines display the results for the strong coupling approximations given by Eqs. (\ref{eq:ma_oblique_strong_kr}), (\ref{eq:ma_oblique_strong_ki}), and (\ref{eq:k3_general}).
    
    For the cases of a strongly ionized plasma (with $\chi = 0.01$) and a partially ionized plasma (with $\chi =1$), depicted in panels $a)$, $b)$, $c)$, and $d)$ of Fig. \ref{fig:oblique}, the behaviors of the fast and the neutral-acoustic modes are similar to those shown in Fig. \ref{fig:perp} for perpendicular propagation. The main difference with respect to Fig. \ref{fig:perp} comes from the inclusion of the slow magneto-acoustic modes, which in these physical conditions are more strongly attenuated than the fast modes. Taking into account the analytical approximations for the damping rates given by Eq. (\ref{eq:ma_oblique_weak_kI}) in the weak coupling regime and by Eqs. (\ref{eq:ki_fast_aa}) and (\ref{eq:ki_slow_ceff}) in the strong coupling regime, the larger attenuation of the slow modes is related to their smaller phase speeds in comparison to the fast modes. In addition, the reason why the lines for the neutral and slow modes overlap in the weak coupling limit is that for a hydrogen plasma and for this particular value of the propagation angle ($\theta = \pi /4$), the relation $c_{\rm{c}} \cos \theta = c_{\rm{n}}$ is fulfilled.

    The weakly ionized scenario with $\chi = 100$, represented in panels $e)$ and $f)$ of Fig. \ref{fig:oblique}, is more complex. Now, the global fast mode is mainly related to the neutral-acoustic mode, as shown by Eq. (\ref{eq:kr_fast_ceff}). On the other hand, the global slow mode has a magnetic nature, since it depends on the Alfvén speed of the plasma as shown by Eq. (\ref{eq:kr_slow_aa}), and can be connected to the fast mode of the charged fluid. Finally, the slow mode in the weak coupling regime tends to the strongly attenuated mode given by Eq. (\ref{eq:k3_general}) as the ratio $\nu / \omega$ increases. Regarding the damping rates, the modes associated with the charged fluid are more strongly damped than those related to the neutral fluid. This can be explained by the fact that $\nu_{\rm{cn}} \gg \nu_{\rm{nc}}$.

    Then, Fig. \ref{fig:parallel} shows the results for waves propagating along the parallel direction to the background magnetic field, that is with $\theta = 0$, and the same plasma parameters as those used for Figs. \ref{fig:perp} and \ref{fig:oblique}. Here, the solutions for the neutral-acoustic mode and the slow mode of the charged fluid no longer overlap in the weak coupling regime. Nevertheless, the wavenumbers represented in the left panels generally resemble those for oblique propagation depicted in Fig. \ref{fig:oblique}. Larger differences appear in the damping rates for the cases with $\chi = 0.01$ and $\chi = 1$, where it can be seen that the slow modes are more strongly attenuated than the fast modes in the weak coupling limit (due to the smaller phase speed of the former) but this trend is inverted as the coupling degree is increased.

    For the particular case of hydrogen plasma, using Eqs. (\ref{eq:ki_fast_aa}) and (\ref{eq:ki_fast_largecc}), the damping rates for Alfvén waves and acoustic waves propagating along the parallel direction can be written in terms of the global Alfvén and sound speeds as
    \begin{equation} \label{eq:ki_alfven}
        k_{\rm{I,Alf}} \approx \frac{\chi \omega^{2}}{2 \nu_{\rm{nc}} \left(1 + \chi \right) a}
    \end{equation}
    and
    \begin{equation} \label{eq:ki_acoustic}
        k_{\rm{I,ac}} \approx \frac{\chi \omega^{2}}{2 \nu_{\rm{nc}} \left(1 + \chi \right) \left(2 + \chi \right)^2 c},
    \end{equation}
    respectively. Then, the ratio between the two damping rates is given by:
    \begin{equation} \label{eq:ratio}
        \mathcal{R_{\rm{k}}} = \frac{k_{\rm{I,Alf}}}{k_{\rm{I,ac}}} \approx \left(2 + \chi \right)^{2}\frac{c}{a},
    \end{equation}
    which shows that for weakly ionized plasmas ($\chi \gg 1$) the damping rate of Alfvén waves is usually larger than that of acoustic waves. However, the opposite can occur for situations with $a \gg c$ and small values of $\chi$. The set of parameters used in the current section provides the following ratios: $\mathcal{R}_{\rm{k}} \left(\chi = 0.01 \right) \approx 1.3$, $\mathcal{R}_{\rm{k}}\left(\chi = 1\right) \approx 3.5$, and $\mathcal{R}_{\rm{k}} \left(\chi = 100 \right) \approx 23500$. These values are in good agreement with the results displayed in the right panels of Fig. \ref{fig:parallel} if we take into account that for $\chi = 0.01$ and $\chi = 1$ the Alfvén wave corresponds to the global fast mode and the acoustic wave is related to the global slow mode, while these relations are swapped for $\chi = 100$.

    The results represented in Figs. \ref{fig:perp} - \ref{fig:parallel} correspond to a particular choice of the relation between the characteristic speeds $c_{\rm{A}}$ and $c_{\rm{c}}$. Therefore, the provided discussion cannot be straightforwardly extrapolated to a different set of physical conditions. Nevertheless, their comparison serves as an example on how the properties of magneto-acoustic waves propagating in a two-fluid partially ionized plasma strongly depend on many parameters, namely the propagation angle, the ionization degree, the strength of the collisional coupling and the relations between the characteristic speeds of each component of the plasma.

    To conclude this section, we note that in contrast to the case of waves generated by an impulsive driver studied in Refs. \onlinecite{Soler2013ApJS..209...16S,Soler2024RSPTA.38230223S,Molevich2024PhPl...31d2115M}, here we do not find cutoff regions where the solutions of the dispersion relation are purely imaginary and waves cannot propagate. This result is in line with those discussed in Ref. \onlinecite{Soler2013ApJ...767..171S} for Alfvén waves. 

\subsubsection{Dependence on the angle of propagation}
    As it has been shown in the previous sections, the approximations for magneto-acoustic waves propagating along the parallel direction to the background magnetic field can be directly obtained from those for propagation at an arbitrary angle. However, if we compare the expressions derived in Sections \ref{sec:ma_weakly_oblique} and \ref{sec:ma_strong_oblique} for oblique propagation with those presented in Section \ref{sec:ma_perp} for perpendicular propagation, it can be seen that it is not straightforward to recover the approximations for the latter case by imposing the value $\theta = \pm \pi/2$ in the more general expressions. The reason is that the general dispersion relation, Eq. (\ref{eq:ma_dr_full}), has a singular point at $\theta = \pm \pi/2$ when solved for $k$. Thus, at that particular value of the propagation angle, we get from Eqs. (\ref{eq:ma_oblique_weak_kr}) and (\ref{eq:ma_oblique_strong_kr}) that $k_{\rm{R}} \to \infty$ for both the fast and slow magneto-acoustic waves, while the approximations derived in Section \ref{sec:ma_perp} result in finite values for the perpendicularly propagating fast waves. Neither the strongly attenuated modes for perpendicular propagation, given by Eq. (\ref{eq:sols_perp_strong2}), are particular cases of the strongly attenuated modes for propagation at an arbitrary angle, given by Eq. (\ref{eq:k3_general}). Therefore, from the analytical point of view, it is convenient to treat in a separate way the cases with $\theta = \pm \pi/2$ and with $\theta \neq \pm \pi/2$, as it has been done in the present work.

\begin{figure}
        \centering
        \includegraphics[width=\hsize]{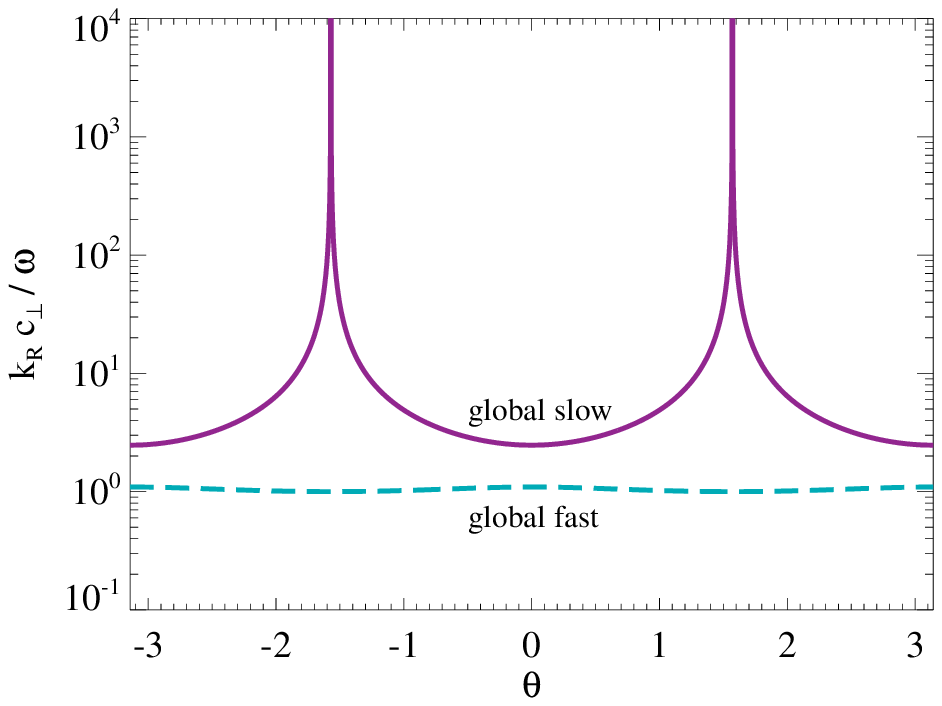}
        \caption{Normalized wavenumber of the ideal MHD magneto-acoustic waves as functions of the angle of propagation. A hydrogen plasma with $c_{\rm{A}}^{2} = 10 c_{\rm{c}}^{2}$ and $\chi = 100$ is considered.}
        \label{fig:kr_theta_ideal}
    \end{figure}

    Nevertheless, it is interesting to investigate this matter from the numerical perspective. Therefore, in this section we study the properties of magneto-acoustic waves as functions of the propagation angle, $\theta$. In the first place, we represent in Fig. \ref{fig:kr_theta_ideal} the results that would correspond to ideal MHD. Since the ideal MHD model is equivalent to assuming that $\{\nu_{\rm{cn}}, \nu_{\rm{nc}} \}\to \infty$ \citep{Soler2013ApJS..209...16S}, these solutions are obtained from Eqs. (\ref{eq:kr4}) or (\ref{eq:kr4_global}). In this case the coupling between the two fluids is perfect and there is no damping due to the collisional interaction, so $k_{\rm{I}} = 0$. Moreover, as shown in Fig. \ref{fig:kr_theta_ideal} there are only two different modes, with the slow mode having a much stronger dependence on the propagation angle than the fast mode: as $\theta \to \pm \pi / 2$, the wavenumber of the slow mode grows without limit, which would lead to a vanishing phase speed, while the wavenumber of the fast mode remains finite.

     \begin{figure*}
        \centering
        \includegraphics[width=0.49\hsize]{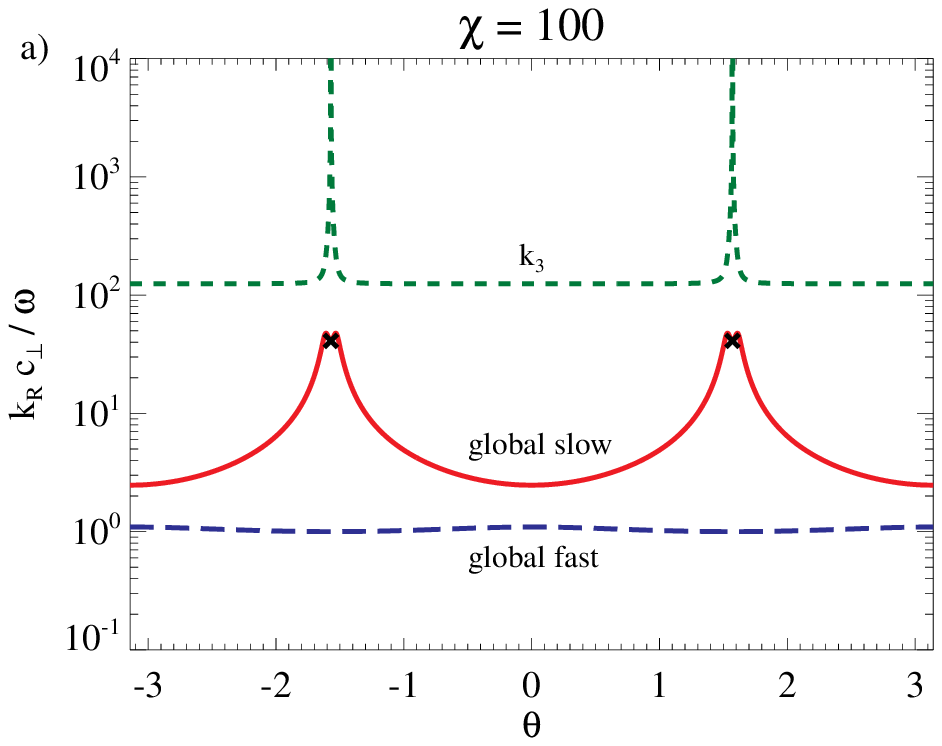}
        \includegraphics[width=0.49\hsize]{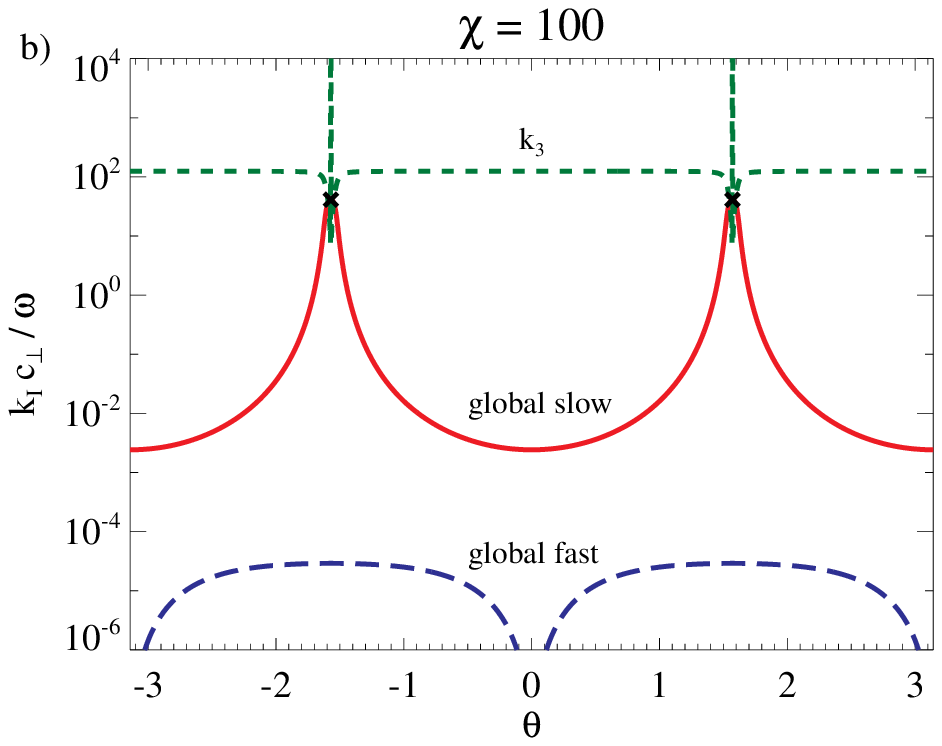}
        \caption{Normalized wavenumbers (panel a) and damping rates (panel b) of 2F magneto-acoustic waves as functions of $\theta$ for a case with a strong collisional coupling given by $\nu / \omega = 10^{3}$. Global fast and slow modes are represented by blue long-dashed and red solid lines, respectively, while the green dashed lines correspond to the mode given by Eq. (\ref{eq:k3_general}). Same plasma parameters as in Fig. \ref{fig:kr_theta_ideal} have been used. The black crosses represent the values of the strongly attenuated mode for perpendicular propagation given by Eq. (\ref{eq:sols_perp_strong2}).}
        \label{fig:2f_theta}
    \end{figure*}

    \begin{figure}
        \centering
        \includegraphics[width=\hsize]{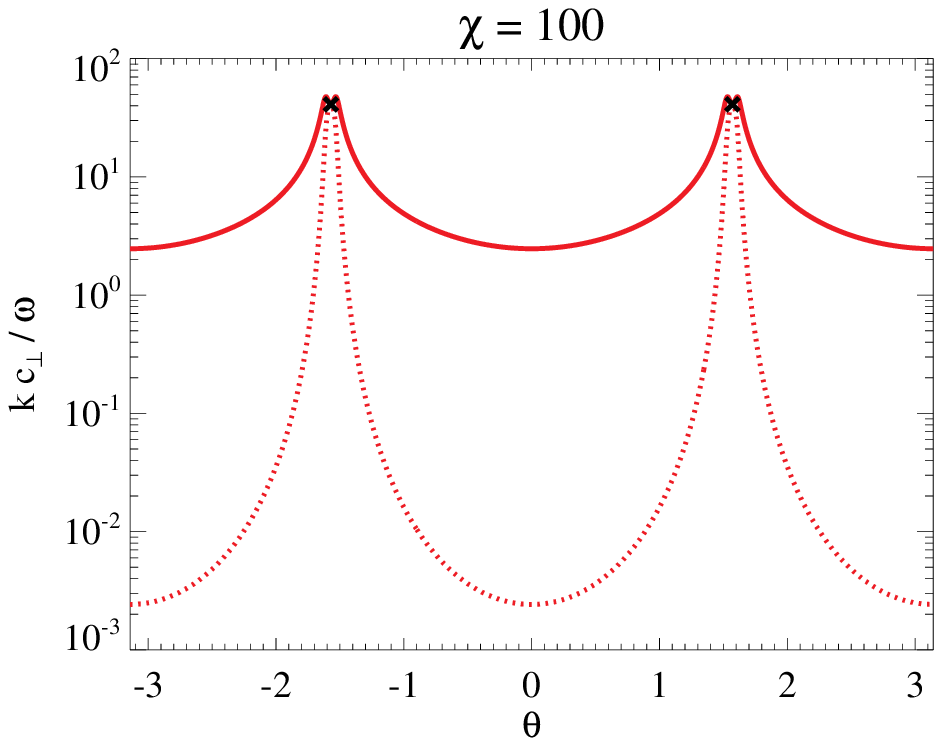}
        \caption{Normalized wavenumber (solid line) and damping rate (dotted line) of the global slow mode represented in Fig. \ref{fig:2f_theta}. The black crosses represent the values of the strongly attenuated mode for perpendicular propagation given by Eq. (\ref{eq:sols_perp_strong2}).}
        \label{fig:slow_theta}
    \end{figure}

    Then, Fig. \ref{fig:2f_theta} shows the exact solutions of the two-fluid dispersion relation, given by Eq. (\ref{eq:ma_dr_full}) for a case of strong but finite collisional coupling, with $\nu / \omega = 10^{3}$. On the one hand, the left panel of this figure shows that the wavenumber of the two-fluid global fast mode has the same behavior as the one from ideal MHD. Regarding the damping rate, the right panel shows that it is maximum for perpendicular propagation and minimum for parallel propagation. On the other hand, the two-fluid global slow mode does not exactly replicate the behavior of the ideal MHD slow mode. For most of the range of propagation angles, their behaviors agree. But as $\theta \to \pm \pi/2$, the wavenumber and the damping rate of the slow mode remain finite, coinciding with the values provided by Eq. (\ref{eq:sols_perp_strong2}) for the strongly attenuated mode in perpendicular propagation. In addition, it is the mode denoted by $k_{\rm{3}}$, given by Eq. (\ref{eq:k3_general}) and which is in general independent from $\theta$, the one that grows without limit as we approach the perpendicular direction of propagation. Therefore, the behavior of the ideal MHD slow mode in the neighborhood of $\theta = \pm \pi/2$ is reproduced in the two-fluid model by a combination of the two-fluid slow mode and the mode that we have denoted by $k_{\rm{3}}$.

    For the sake of simplicity, in Fig. \ref{fig:2f_theta} we have only represented the case with a density ratio of $\chi = 100$, but similar conclusions can be extracted for other cases, with smaller values of $\chi$ showing a better agreement between the two-fluid and the ideal MHD results. In addition, if we increase the ratio $\nu / \omega$, the agreement between the two-fluid and ideal MHD slow modes increases and the mode denoted by $k_{\rm{3}}$ is damped by collisions at even shorter scales. Nevertheless, the discussion from the previous paragraph still applies.

    Finally, Fig. \ref{fig:slow_theta} displays a comparison between the wavenumber and the damping rate of the global slow mode represented in Fig. \ref{fig:2f_theta}. This plot allows us to understand why some of the approximations derived in Sections \ref{sec:ma_weakly_oblique} and \ref{sec:ma_strong_oblique} for a general value of $\theta$ do not apply to the particular case of $\theta = \pm \pi/2$. Those approximations resulted from the assumption that the damping rates are much smaller than the wavenumber, $k_{\rm{I}} \ll k_{\rm{R}}$. As it can be checked from Fig. \ref{fig:2f_theta}, this condition is fulfilled for the global fast mode for any value of the propagation angle. On the contrary, Fig. \ref{fig:slow_theta} shows that the wavenumber and the damping rate of the slow mode tend to become identical at $\theta = \pm \pi/2$, so the condition $k_{\rm{I}} \ll k_{\rm{R}}$ is no longer fulfilled, breaking the justification for the use of the respective approximations.

\section{Discussion} \label{sec:discussion}
    After the detailed study of the solutions of the dispersion relation performed in the previous section, here we discuss some general trends for the limits of weak and strong collisional coupling, comparing the behaviors of impulsive driven waves described in previous works with those for periodically driven waves obtained in the present investigation.

\subsection{Weak coupling}
    As shown in Refs. \onlinecite{Mouschovias2011MNRAS.415.1751M,Soler2013ApJ...767..171S,Soler2024RSPTA.38230223S,Molevich2024PhPl...31d2115M}, in the weak coupling regime the frequencies and damping rates of waves generated by an impulsive driver with a real $k = k_{\rm{R}}$ can be generally written as
    \begin{equation} \label{eq:w_weak_gen}
        \omega_{\rm{R}} \approx \pm \mathcal{C}_{\rm{\alpha}} k_{\rm{R}} \quad \text{and} \quad \omega_{\rm{I}} \approx -\frac{\nu_{\rm{\alpha \beta}}}{2}, 
    \end{equation}
    where $\mathcal{C}_{\rm{\alpha}}$ is the phase speed of a particular wave mode associated with the species $\alpha$ (for instance, $\mathcal{C}_{\rm{\alpha}} = c_{\rm{n}}$ for the acoustic wave of the neutral fluid, or $\mathcal{C}_{\rm{\alpha}} = c_{\rm{A}}$ for Alfvén waves in the charged fluid) and $\nu_{\rm{\alpha \beta}}$ is the collision frequency of species $\alpha$ with particles from species $\beta$. According to this expression, for a fixed value of $k_{\rm{R}}$ the frequency of a wave increases with its phase speed (which, depending on the nature of the wave, grows with the magnetic field strength and/or with the temperature), but its damping rate is independent from the wavenumber. This implies that all the modes associated with a given species $\alpha$ have the same damping rate. For instance, the Alfvén mode and the slow and fast magneto-acoustic modes of the charged fluid have the same damping rate although their frequencies can be very different.

    For the case of periodically driven waves with a real $\omega = \omega_{\rm{R}}$, the results from Section \ref{sec:results} show that the wavenumbers and spatial damping rates can be typically written in the form
    \begin{equation} \label{eq:k_weak_gen}
        k_{\rm{R}} \approx \pm \frac{\omega_{\rm{R}}}{\mathcal{C}_{\rm{\alpha}}} \quad \text{and} \quad k_{\rm{I}} \approx \pm \frac{\nu_{\rm{\alpha \beta}}}{2 \mathcal{C}_{\rm{\alpha}}}.
    \end{equation}
    Therefore, both parameters decrease as the wave phase speed increases and now slower waves have larger damping rates than the faster ones. This is a clear departure from the trend usually found for impulsively generated waves. Thus, the behavior of the temporal damping rates ($\omega_{\rm{I}}$) and the spatial damping rates ($k_{\rm{I}}$) are not equivalent, although, as it is shown below, they are related.

    A common procedure to measure the relative importance of the damping of the waves is to compute their quality factor, which is usually defined as
    \begin{equation} \label{eq:qw}
        Q_{\rm{\omega}} = \frac{1}{2}\frac{|\omega_{\rm{R}}|}{|\omega_{\rm{I}}|}
    \end{equation}
    for standing waves, and as
    \begin{equation} \label{eq:qk}
        Q_{\rm{k}} = \frac{1}{2}\frac{|k_{\rm{R}}|}{|k_{\rm{I}}|}
    \end{equation}
    for waves generated by a periodic driver. Inserting Eqs. (\ref{eq:w_weak_gen}) and (\ref{eq:k_weak_gen}) into the definitions of the quality factors, we find that
    \begin{equation} \label{eq:qw_qk}
        Q_{\rm{\omega}} = \frac{\mathcal{C}_{\rm{\alpha}} k_{\rm{R}}}{\nu_{\rm{\alpha \beta}}} \quad \text{and} \quad Q_{\rm{k}} = \frac{\omega_{\rm{R}}}{\nu_{\rm{\alpha \beta}}}.
    \end{equation}
    Then, if we take into account that the real parts of the wavenumber and the wave frequency are related by $\omega_{\rm{R}} = \mathcal{C}_{\rm{\alpha}} k_{\rm{R}}$, we see that the two quality factors have the same value. Therefore, we can use a common parameter $Q_{\rm{D}}= Q_{\rm{\omega}} = Q_{\rm{k}}$ to describe the damping due to ion-neutral collisions for both types of driving mechanisms. A similar relation was found in Ref. \onlinecite{Terradas2010A&A...524A..23T} in the study of resonant absorption of kink waves, which demonstrated that the damping per wavelength is exactly the same as the damping per period as long as the weak damping approximation remains valid and the expressions given by Eq. (\ref{eq:ki_wi_tagger}), which are the same as Eqs. (A3) and (A4) from Ref. \onlinecite{Tagger1995A&A...299..940T}, can be applied.

    The fact that the damping rates given by Eqs. (\ref{eq:w_weak_gen}) and (\ref{eq:k_weak_gen}) do not depend on the wavenumber or the frequency of the waves can be understood as follows. In the weak coupling limit, we have that $\omega_{\rm{R}} \gg \nu_{\rm{\alpha \beta}}$ and the oscillation frequency does not depend either on the collision frequency, meaning that the properties of a MHD wave in a given component of the plasma are mainly determined by the magnetic and pressure forces that directly affect that component, while the collisional interaction with the other fluid plays a secondary role. The oscillation frequency of the wave can vary due to the strength of the restoring forces acting on one individual fluid, but the interaction with the other component of the plasma occurs at a fixed rate determined by the collision frequency. Therefore, all the waves within the range of validity of the weak coupling approximation will be affected by the inter-species collisions at the same fixed rate. This can also explain why the spatial damping rates for periodically driven waves are inversely proportional to the phase speed: faster waves traverse a larger distance before the collisional interaction takes place; therefore, their damping length (which is inversely proportional to $k_{\rm{I}}$) is larger.
    
\subsection{Strong coupling}
    In Sections \ref{sec:ma_perp_coupled} and \ref{sec:ma_strong_oblique} it has been shown that in the strong coupling limit the spatial damping rates of fast waves propagating in the perpendicular direction to the magnetic field are given by Eq. (\ref{eq:perp_strong_kId1}) and those of magneto-acoustic waves propagating in the parallel direction are given by Eqs. (\ref{eq:ki_fast_aa}) and (\ref{eq:ki_fast_largecc}). After some manipulation, these expressions can be written in a more compact form, namely
    \begin{equation} \label{eq:ki_strong_gen}
        k_{\rm{I}} = \pm \frac{\chi \omega^{2}}{2 \nu_{\rm{nc}} \left(1 + \chi \right)^{3}} \frac{ \left(\mathcal{C}_{\rm{\alpha}}^{2} - \mathcal{C}_{\rm{\beta}}^{2} \right)^{2}}{\mathcal{C}_{\rm{T}}^{5}},
    \end{equation}
    where $\mathcal{C}_{\rm{\alpha}}$ and $\mathcal{C}_{\rm{\beta}}$ are the phase speeds of the wave modes associated with each individual fluid and $\mathcal{C}_{\rm{T}}$ is the phase speed of the global wave mode. For the case of perpendicular propagation, the set of phase speeds is given by
    \begin{equation} \label{eq:c_perp}
        \mathcal{C}_{\rm{\alpha}} = \sqrt{c_{\rm{A}}^{2} + c_{\rm{c}}^{2}}, \quad \mathcal{C}_{\rm{\beta}} = c_{\rm{n}},
    \end{equation}
    and
    \begin{equation} \label{eq:cT_perp}
        \mathcal{C}_{\rm{T}} = \sqrt{\frac{c_{\rm{A}}^{2}+c_{\rm{c}}^{2}+\chi c_{\rm{n}}^{2}}{1+\chi}}=\sqrt{a^{2} + c^{2}}.
    \end{equation}
    The corresponding parameters for acoustic waves is obtained from Eqs. (\ref{eq:c_perp}) and (\ref{eq:cT_perp}) by setting $c_{\rm{A}} = 0$ and $a = 0$, while the case of Alfvén waves is represented by
    \begin{equation} \label{eq:c_alfven}
        \mathcal{C}_{\rm{\alpha}} = c_{\rm{A}}, \quad \mathcal{C}_{\rm{\beta}} = 0, \quad \text{and} \quad \mathcal{C}_{\rm{T}} = a.
    \end{equation}

    In a similar way, using Eq. (\ref{eq:perp_strong_wId1}) as reference, the following expression for the damping rates of waves generated by an impulsive driver can be obtained:
    \begin{equation} \label{eq:wi_strong_gen}
        \omega_{\rm{I}} = -\frac{\chi k^{2}}{2 \nu_{\rm{nc}} \left(1 + \chi \right)^{3}} \frac{ \left(\mathcal{C}_{\rm{\alpha}}^{2} - \mathcal{C}_{\rm{\beta}}^{2} \right)^{2}}{\mathcal{C}_{\rm{T}}^{2}}.
    \end{equation}

    Equations (\ref{eq:ki_strong_gen}) and (\ref{eq:wi_strong_gen}) show that the damping rates grow with the difference of the characteristic speeds of each separate fluid and decrease as the global characteristic speed is increased. In addition, they can be used to explain why in the strong coupling regime magnetic waves are typically more affected than acoustic waves by the ion-neutral damping \citep{Forteza2007A&A...461..731F,Soler2015ApJ...810..146S,CallyGomezMiguez2023ApJ...946..108C}. On the one hand, since collisions tend to balance the temperatures of the components of the plasma \citep{Spitzer1956pfig.book.....S}, the different sound speeds tend to be of the same order of magnitude (unless there are large differences in the mass of the particles that conform each fluid). Therefore, the three characteristic speeds used in Eqs. (\ref{eq:ki_strong_gen}) and (\ref{eq:wi_strong_gen}) fulfill that $\mathcal{C}_{\rm{\alpha}} \sim \mathcal{C}_{\rm{\beta}} \sim \mathcal{C}_{\rm{T}}$. On the other hand, for the case of Alfvén waves, the relations between those three parameters strongly depend on the ionization degree of the plasma, as shown by Eq. (\ref{eq:c_alfven}), and they can vary by several orders of magnitude.

    Focusing on the scenario in which the magnetic field dominates the dynamics of the plasma, Eq. (\ref{eq:ki_strong_gen}) reduces to Eq. (\ref{eq:ki_alfven}), while Eq. (\ref{eq:wi_strong_gen}) reduces to
    \begin{equation} \label{eq:wi_alf_strong}
        \omega_{\rm{I,Alf}} = -\frac{\chi k^{2}a^{2}}{2 \nu_{\rm{nc}} \left(1 + \chi \right)}, 
    \end{equation}
    which is equivalent to the expressions derived in Refs. \onlinecite{Braginskii1965RvPP....1..205B,Forteza2007A&A...461..731F}.

    It can be seen from Eqs. (\ref{eq:ki_alfven}) and (\ref{eq:wi_alf_strong}) that increasing the magnetic field leads to a smaller $k_{\rm{I}}$ but a larger $\omega_{\rm{I}}$. The former result is explained by the growth of the wavelength as the phase speed is increased: the strong coupling regime occurs when the wavelength is much larger than the collisional mean free path, so a larger wavelength produces an even stronger coupling between the species and, consequently, a reduced damping rate. In contrast, a larger phase speed leads to a larger oscillation frequency, which results in a growth of the damping rate $\omega_{\rm{I}}$ because the fluids become more weakly coupled as the oscillation frequency approaches the collision frequency.

    From Eqs. (\ref{eq:ki_strong_gen}) and (\ref{eq:wi_strong_gen}), and taking into account that in the strong coupling limit $\omega_{\rm{R}} = \mathcal{C}_{\rm{T}} k_{\rm{R}}$, the following expression for the quality factor can be derived:
    \begin{equation} \label{eq:q_strong}
        Q_{\rm{D}} = \frac{\nu_{\rm{nc}} \left(1 + \chi \right)^{3} \mathcal{C}_{\rm{T}}^{4}}{\chi \omega_{\rm{R}} \left(\mathcal{C}_{\rm{\alpha}}^{2} - \mathcal{C}_{\rm{\beta}}^{2} \right)^{2}}.
    \end{equation}
    Thus, the quality factor for Alfvén waves is given by
    \begin{equation} \label{eq:q_strong_alfven}
        Q_{\rm{D,Alf}} = \frac{\nu_{\rm{nc}} \left(1 + \chi \right)}{\chi \omega_{\rm{R}}},
    \end{equation}
    and the corresponding parameter for acoustic waves in a hydrogen plasma is given by
    \begin{equation} \label{eq:q_strong_ac}
        Q_{\rm{D,ac}} = \frac{\nu_{\rm{nc}} \left(1 + \chi \right) \left(2 + \chi \right)^{2}}{\chi \omega_{\rm{R}}}.
    \end{equation}
    The comparison between Eqs. (\ref{eq:q_strong_alfven}) and (\ref{eq:q_strong_ac}) shows that $Q_{\rm{D,Alf}} < Q_{\rm{D,ac}}$, which means that the relative importance of the damping due to ion-neutral collisions is larger for magnetic waves than for acoustic waves, in agreement with the discussion in Ref. \onlinecite{CallyGomezMiguez2023ApJ...946..108C}.

    Finally, we note that the expressions discussed in the present section do not generally apply to the propagation of magneto-acoustic waves at an arbitrary angle $\theta$, since the interaction between the charged and neutral fluids leads to more convoluted expressions for the wavenumbers and the damping rates, as shown by Eqs. (\ref{eq:ma_oblique_strong_kr}) and (\ref{eq:ma_oblique_strong_ki}). However, they provide a useful reference to understand the overall dependence of the damping rates on the parameters of the plasma.
    
\section{Summary and conclussions} \label{sec:summary}
    In this work, we used a two-fluid plasma model \citep[see, e.g.,][]{Zaqarashvili2011A&A...529A..82Z,Soler2013ApJ...767..171S,Soler2013ApJS..209...16S,Khomenko2014PhPl...21i2901K} to perform a detailed study of the properties of magneto-acoustic waves in partially ionized plasmas. Mainly focusing on the case of propagating waves, we analyzed their wavenumbers and damping rates for a wide range of parameters, such as different ionization degrees, propagation angles, or strengths of the collisional coupling between the charged and neutral components of the plasma. 

    We described how the different modes are affected by the physical conditions of the plasma and how they are related to the characteristic properties of each individual fluid. For instance, in the weak coupling regime, each solution of the dispersion relation can be directly connected to the fast and slow magneto-acoustic modes of the charged fluid or to the acoustic mode of the neutral fluid. However, in the strong coupling regime the properties of the individual modes mix to give rise to the global fast and slow magneto-acoustic waves, in line with the discussion performed in Ref. \onlinecite{Soler2013ApJS..209...16S} for the case of standing waves, and to an additional type of mode that is strongly attenuated by the effect of collisions. In the investigation performed in Ref. \onlinecite{CallyGomezMiguez2023ApJ...946..108C}, this additional mode was associated with the neutral-acoustic wave only, but here we show that it generally depends on the properties of both fluids. 

    Furthermore, we derived analytical approximations for both the regimes of weak and strong collisional coupling, and checked their good agreement with the exact numerical solutions from the full dispersion relation. When the collision frequencies are much smaller than the wave frequency, the analytical approximations show that the damping rates due to collisions are directly proportional to the collision frequencies but inversely proportional to the phase speed of the wave. In the opposite limit, the damping rates are inversely proportional to the collision frequencies and for some particular configurations, such as fast waves propagating along the perpendicular direction to the background magnetic field or for acoustic and Alfvén waves, they can be written in terms of the difference between the characteristic speeds of each fluid. In addition, we performed a comparison between the general trends found in the present work for periodically driven waves and those for impulsively excited waves described in previous studies \citep{Braginskii1965RvPP....1..205B,Forteza2007A&A...461..731F,Mouschovias2011MNRAS.415.1751M,Soler2024RSPTA.38230223S,Molevich2024PhPl...31d2115M}, showing how the dependence of the damping rates on the phase speed of the waves varies between the two kinds of driving mechanisms.

    We also paid particular attention to the dependence of the solutions of the dispersion relation on the angle of propagation. In this way, we found that the approximate expressions for propagation at an arbitrary angle cannot be straightforwardly applied to the case of perpendicular propagation. The reason is that the dispersion relation, Eq. (\ref{eq:ma_dr_full}), has a singular point at $\theta = \pm \pi/2$ when solved for a real wave frequency $\omega$. Moreover, it is found that the condition used to obtain the approximations, which is the damping rates being much smaller than the wavenumbers ($k_{\rm{I}} \ll k_{\rm{R}}$), is not fulfilled for slow magneto-acoustic waves. Therefore, it is convenient to treat the case of perpendicular propagation separately from the more general one.

    Here, we focused only on the effect that elastic collisions between charges and neutrals have on the properties of magneto-acoustic waves. However, there are other non-ideal processes that may also have a strong impact. For instance, it was shown in Refs. \onlinecite{Amagishi1993PhRvL..71..360A,Pandey2008MNRAS.385.2269P,Pandey2015MNRAS.447.3604P,MartinezGomez2025ApJ...982....4M} that the inclusion of Hall's current in the model becomes fundamental for the description of Alfvén waves in weakly ionized plasmas. Refs. \onlinecite{Terradas2001A&A...378..635T,Carbonell2004A&A...415..739C,Terradas2005A&A...434..741T,Forteza2008A&A...492..223F,Barcelo2011A&A...525A..60B,Soler2015ApJ...810..146S,PandeyWardle2024MNRAS.535.3410P} demonstrated by means of single-fluid models that magneto-acoustic waves may be also efficiently damped by radiative losses, viscosity or thermal conduction. And Ref. \onlinecite{Ballai2019FrASS...6...39B} showed that the processes of ionization and recombination introduce an additional damping mechanism for Alfvén and slow waves in partially ionized plasmas. Furthermore, we have considered that the different wave modes are independent solutions of the dispersion relation that do not interact with each other. However, there are processes such as parametric resonances \citep{Shergelashvili2005A&A...429..767S,Shergelashvili2007PhRvE..76d6404S}, resonant absorption \citep{Hollweg1990ApJ...349..335H,Terradas2008ApJ...679.1611T}, mode conversion \citep{Schunker2006MNRAS.372..551S,Cally2019ApJ...885...58C} or the presence of shear flows \citep{Chagelishvili1994PhRvE..50.4283C,Cantallops2025A&A...701A.238C} which produce a coupling between the MHD modes and can result in damping or amplification of the waves. Therefore, it would be interesting to include these effects and interactions in future studies to obtain a better description of magneto-acoustic waves propagating in two-fluid partially ionized plasmas.

    Finally, one of the main assumptions of the present study is that small-amplitude perturbations are applied to a uniform background plasma. This leads to a symmetry between the solutions of the dispersion relation corresponding to forward and backward propagating waves. This symmetry is broken when non-uniform backgrounds are considered, as shown for instance in Refs. \onlinecite{Popescu2019A&A...627A..25P,Popescu2019A&A...630A..79P} for the case of stratified atmospheres. Nevertheless, the results discussed here are good approximations for the damping rates due to elastic collisions when the wavelengths are smaller than the spatial scales in which the non-uniform background varies. In addition, the evolution of the temperature of the plasma due to the charge-neutral frictional heating \citep[see, e.g.,][]{Piddington1956MNRAS.116..314P,Osterbrock1961ApJ...134..347O, Leake2005A&A...442.1091L,Song2011JGRA..116.9104S,Khomenko2012ApJ...747...87K,Arber2016ApJ...817...94A} or the influence of collisions in the properties of shocks \citep[see, e.g.,][]{Hillier2016A&A...591A.112H,Snow2021A&A...645A..81S} and in the process of wave steepening \citep{MartinezGomez2018ApJ...856...16M,Popescu2019A&A...630A..79P,Ballester2020A&A...641A..48B,Zhang2021ApJ...911..119Z} cannot be captured by the linear analytical approach used here. A non-linear treatment would be required, which is left for future works.

\begin{acknowledgements}
    D.M. acknowledges financial support from \textit{Govern de les Illes Balears} and \textit{Fons Social Europeu Plus (FSE+)} through the grant PD/014/2023 of the \textit{Vicenç Mut} program. This publication is part of the R+D+i project PID2023-147708NB-I00, funded by MCIN/AEI/10.13039/501100011033 and by FEDER, EU. D.M. also thanks Roberto Soler for his comments on an early draft of this paper and for providing very useful references, and the anonymous referees for their helpful suggestions and remarks.
\end{acknowledgements}

\bibliography{ma_waves}

\end{document}